\def\epsfpreprint{Y}   
                       
                       \if \epsfpreprint Y
\documentstyle[12pt,epsf]{article} \fi \if \epsfpreprint N
\documentstyle[12pt]{article} \fi

\def\figure#1#2#3{\if \epsfpreprint Y \epsfxsize=#3 truein
\centerline{\epsffile{fig_#1.eps}}
\centerline{\vbox{{\bf \noindent Figure #1.} #2}}
\bigskip \fi}

\def\Psibar{\overline{\Psi}}

\def\chibar{\overline{\chi}}
\def\chibarchi{\langle \overline{\chi} \chi \rangle}
\def\chibarchim{\overline{\chi} \chi}
\def\phibar{\overline{\phi}}

\def\slash{\!\!\!\!/}

\def\meff{m_{\rm eff}}
\def\mres{m_{\rm res}}

\def\a0limit{a \rightarrow 0}
\def\cdof{\chi^2 / {\rm dof} }

\def\spose#1{\hbox to 0pt{#1\hss}}
\def\ltapprox{\mathrel{\spose{\lower 3pt\hbox{$\mathchar"218$}}
 \raise 2.0pt\hbox{$\mathchar"13C$}}}
\def\gtapprox{\mathrel{\spose{\lower 3pt\hbox{$\mathchar"218$}}
 \raise 2.0pt\hbox{$\mathchar"13E$}}}
\def\inapprox{\mathrel{\spose{\lower 3pt\hbox{$\mathchar"218$}}
 \raise 2.0pt\hbox{$\mathchar"232$}}}

\def\figsizea{6.4}

\def\one{The magnitude of the Wilson line $<|W|>$ in an $8^4$
lattice. The diamonds are from the quenched theory. The cross is
from the dynamical theory at $\beta=2.3$ with $L_s=24$, $m_f=0.0$
and $m_0=1.9$.
}
\def\two{The gluino condensate values generated by the computer
starting from an ordered initial configuration (solid line) and from a
dis-ordered initial configuration (dotted line).  The x axis is the
configuration number and corresponds to ``computer time''. This is
from a simulation of the full theory on an $8^4$ lattice at
$\beta=2.3$ with $L_s=12$, $m_f=0.04$ and $m_0=1.9$.
}
\def\three{The gluino condensate vs. $m_f$ for various values of $L_s$
from the dynamical theory on an $8^4$ lattice at $\beta=2.3$ and
$m_0=1.9$. The fits are to the function in eq. \ref{linear}.
}
\def\four{The extrapolated values of the fits in fig. 3 vs. $L_s$.
The fit is to the function in eq. \ref{exponential}.  }
\def\five{The gluino condensate vs. $L_s$ for various  values
of $m_f$ from the dynamical theory on an $8^4$ lattice at $\beta=2.3$
and $m_0=1.9$. The fits are to the function in eq. \ref{exponential}.
}
\def\six{The extrapolated values of the fits in fig. 5 vs. $m_f$.
The fit is to the function in eq. \ref{linear}.
}
\def\seven{The gluino condensate vs. $L_s$ from the dynamical theory
on an $8^4$ lattice at $\beta=2.3$, $m_f=0.0$ and $m_0=1.9$. The fit
is to the function in eq. \ref{exponential}.
}
\def\eight{The gluino condensate vs. $L_s$ from the dynamical theory
on a $4^4$ lattice at $\beta=2.1$, $m_f=0.0$ and $m_0=1.9$ (crosses).
The $8^4$ data of fig. 7 are also plotted for comparison (diamonds).
The fits are to the function in eq. \ref{exponential}.
}
\def\nine{The gluino condensate values generated by the computer
for the full theory on a $4^4$ lattice at $\beta=2.1$ with $m_f=0.0$,
$m_0=1.9$, and for various values of $L_s$.  The x axis is the
configuration number and corresponds to ``computer time''.
}
\def\ten{The histogram of the values of the gluino condensate
from fig. 9 (solid lines). The dotted lines are from an identical set
of simulations, except that the fermion determinant has been set to one
(quenched theory). The time evolutions for the quenched simulations
are not shown in this paper. The area under the curves is normalized
to one.  All sets have 300 data points.
}
\def\eleven{The gluino condensate values generated by the computer
for the full theory on an $8^4$ lattice at $\beta=2.3$ with $m_f=0.0$,
$m_0=1.9$ and $L_s=24$.  The x axis is the
configuration number and corresponds to ``computer time''.
}
\begin{document}

\title{\bf Super Yang-Mills on the lattice with domain wall fermions} 
\vskip 1. truein
\author{George~T.~Fleming$^a$, John~B.~Kogut$^b$ and  Pavlos~M.~Vranas$^b$ \\
\\
\\
${}^a$Physics Department, Ohio State University Columbus, OH 43210 \\
${}^b$Physics Department, University of Illinois, Urbana, IL 61801 \\
\\
\\
}

\maketitle

\begin{abstract}

The dynamical ${\cal N}=1$, SU(2) Super Yang-Mills theory is studied
on the lattice using a new lattice fermion regulator, domain wall
fermions.  This formulation even at non-zero lattice spacing does not
require fine-tuning, has improved chiral properties and can produce
topological zero-mode phenomena. Numerical simulations of the full
theory on lattices with the topology of a torus indicate the formation
of a gluino condensate which is sustained at the chiral limit. The
condensate is non-zero even for small volume and small supersymmetry
breaking mass where zero mode effects due to gauge fields with
fractional topological charge appear to play a role.

\end{abstract}

\newpage

\section{Introduction}
\label{sec_intro}

It is believed that super-symmetric (SUSY) field theories may play an
important role in describing the physics beyond the Standard Model.
Non-perturbative studies of these theories are of great interest.
First-principles numerical simulations may be able to provide
additional information and confirmation of existing analytical
calculations.  Typically first principles numerical simulations of
field theories are done within the framework of the lattice
regulator. A host of results have been produced in this way for many
field theories and most notably QCD. Several SUSY theories can
also be formulated on the lattice and be studied numerically.
To be more specific consider the problems of putting a SUSY
theory on the lattice (see for example 
\cite{Curci_Veneziano,Neuberger_fermions,Kaplan_Schmaltz}):

1) Since space-time is discrete only a subgroup of the Poincar\'{e}
group survives and as a result SUSY is broken.  This problem is not
severe and is of the same nature as in QCD.  The symmetry breaking
operators that are allowed by the remaining symmetries are irrelevant.
One can calculate at several lattice spacings $a$ and then take the
$a \rightarrow 0$ limit. No fine tuning is needed.

2) If the SUSY theory under consideration involves scalar fields one
can have scalar mass terms that break SUSY since typically they are not
forbidden by some symmetry. Since these operators are relevant
fine tuning will be needed in order to cancel their contributions.
The four-dimensional ${\cal N}=1$ Super Yang-Mills
(SYM) theory does not involve scalars and therefore it does not have
this problem.

3) A naive regularization of fermions results in the well known
doubling problem \cite{Nielsen_Ninomiya}.  For each fermion species in
the four-dimensional continuum 16 are generated on the lattice with
total chirality of zero. This results in the wrong number of degrees
of freedom and therefore breaks SUSY. However,
this problem may be possible to treat as in QCD. This is the case
for ${\cal N}=1$ SYM. 

One possible way to remove the unwanted fermion
degrees of freedom is to add an irrelevant operator (Wilson term
\cite{Wilson_fermions}) that gives them heavy masses of the size of the
cutoff. This term unavoidably breaks the chiral symmetry
\cite{Nielsen_Ninomiya} and as a result a gluino mass term is no
longer forbidden. Since such a term is relevant, fine tuning of the
bare fermion mass is necessary as the continuum limit is approached in
order to cancel its contribution. Although fine tuning is not a
welcomed property this method makes it possible to recover the
continuum target theory.

Therefore, it is possible to simulate numerically the ${\cal N}=1$ SYM theory
using existing lattice ``technology'' since all three difficulties can
be circumvented. This observation was made some time ago
\cite{Curci_Veneziano}.  In particular, it was argued that, using a
standard lattice gauge theory action with a pure gauge Wilson
plaquette term and Wilson fermions in the adjoint representation,
numerical simulations could be done. 
Pioneering work using these methods has already produced very
interesting numerical results \cite{Montvay,Donini}.  Also, for
proposed lattice tests of SYM see \cite{Evans}.  For a supersymmetric
formulation on the lattice using Kogut--Susskind \cite{Kogut_Susskind}
fermions see \cite{Banks}.

There are two unwelcomed difficulties in using Wilson fermions. The
first has already been mentioned and it is the need for fine
tuning. The second is of a technical nature. It turns out that the
Pfaffian resulting from the fermionic integration is not positive
definite \cite{Montvay} at finite lattice spacing. However, it does
become positive definite as the continuum limit is approached and
therefore as a ``cure'' only the absolute value of the Pfaffian is
used \cite{Montvay,Donini}. However, this introduces non-analyticities
that may make the approach to the continuum limit difficult.

Both of these difficulties can be brought under control by using an
alternative fermion lattice regulator, domain wall fermions (DWF).
The use of DWF in
supersymmetric theories has been explored in the very nice work of
\cite{Neuberger_fermions,Kaplan_Schmaltz}. 
The methods in this paper are along the lines of these references.
Domain wall fermions were introduced in \cite{Kaplan}, were further
developed in \cite{NN1} and in \cite{Shamir,Furman_Shamir}. They provide 
a new way for treating the
unwelcomed chiral symmetry breaking that is introduced
when the fermion doubler species are removed. Here a variant of this
approach will be used \cite{Shamir,Furman_Shamir}.  For reviews on the
subject please see \cite{DWF_reviews} and references therein. DWF have
already been used for numerical simulations of the two flavor
dynamical Schwinger model \cite{PMV_Schwinger}, 
dynamical QCD
\cite{DWF_dyn_QCD_Columbia}, quenched QCD
\cite{DWF_quenched_QCD_Columbia,DWF_quenched_QCD_RBC,DWF_zero_modes_Columbia,DWF_quenched_QCD_BNL,DWF_quenched_QCD_Japan,DWF_quenched_QCD_Jlab}, 
as well as for simulations of 
4-Fermi models \cite{VTK_four_Fermi}. 
The use of DWF in
supersymmetric theories has also been explored in a different fashion in
\cite{Nishimura,Aoki_Nagai_Zenkin}. 
Furthermore, the use of
overlap \cite{NN1} type fermions has been explored in 
\cite{NN1,Overlap_SUSY,EHN_adjoint_index},
and the use of other related types of fermions has been explored in
\cite{Block_SUSY,GW_SUSY}.

In the lattice DWF formulation of a vector-like theory the fermionic
fields are defined on a five dimensional space-time lattice using a
local action. The fifth direction can be thought of as an extra
space-time dimension or as a new internal flavor space. The gauge
fields are introduced in the standard way in the four dimensional
space-time and are coupled to the extra fermion degree of freedom in a
diagonal fashion. The key ingredient is that the boundary conditions
of the Dirac operator along the fifth direction are taken to be free.
As a result, although all fermions are heavy, two chiral,
exponentially bound surface states appear on the boundaries (domain
walls) with the plus chirality localized on one wall and the minus
chirality on the other.  The two chiralities mix only by an amount that is 
exponentially small in $L_s$, where $L_s$ is the number of lattice sites
along the fifth direction, and form a Dirac spinor that propagates in
the four-dimensional space-time with an exponentially small mass.
Therefore, the amount of chiral symmetry breaking that is artificially
induced by the regulator can be controlled by the new parameter $L_s$.
In the $L_s \to \infty$ limit the chiral symmetry is exact, even at
finite lattice spacing, so there is no need for fine-tuning.

For the first time the approach to the chiral limit has been separated
from the approach to the continuum limit. Furthermore, the computing
requirement is linear in $L_s$.  This is to be contrasted with
traditional lattice fermion regulators where the chiral limit is
approached only as the continuum limit is taken, a process that is
achieved at a large computing cost. Specifically, because of
algorithmic reasons, the computing cost to reduce the lattice spacing
by a factor of two grows by a factor of $2^{8-10}$ in four
dimensions. Therefore, the unique properties of DWF provide a way to
bring under control the systematic chiral symmetry breaking effects
using today's supercomputers.

The purpose of this paper is two-fold. First, the techniques for
performing a numerical simulation of the full ${\cal N}=1$ SU(2) SYM theory
using DWF are collected and it is demonstrated that they work as
expected by performing numerical simulations of the full theory.
Second, the gluino condensate is measured.  It is expected that a
non-zero gluino condensate must form 
\cite{SYM_gluino_cond,Cohen_Gomez,Schaefer,Mattis_etal}.  However,
there are also arguments that the theory has a phase where a gluino
condensate does not form \cite{SYM_without_gluino_cond}.  In the
numerical simulations performed here it is found that a non-zero
gluino condensate is sustained in the limit of zero gluino mass. 
This result is at a finite lattice spacing and therefore SUSY is still
broken albeit by irrelevant operators. 

It must be emphasized that in
this work due to limited computer resources no attempt has been made
to extrapolate to the continuum limit. It is possible that in this
limit the gluino condensate may vanish.  Future work using larger
computer resources could calculate the gluino condensate at several
lattice spacings and extract the continuum value. But even then, it
will never be possible to numerically prove that the finite lattice
spacing theory is not separated from the continuum theory by a phase
transition.  This problem is not particular to the case at hand but is
in the nature of numerical investigations. They can provide strong
evidence but not unquestionable proof. A well known case with similar
problems relates to the question of confinement and chiral symmetry
breaking in QCD.

This paper is organized as follows. In section
\ref{sec_lattice_formulation}, the DWF lattice formulation 
of ${\cal N}=1$ SU(2) SYM
is presented. In section \ref{sec_analy}, analytical considerations
relating to the gluino mass, the Ward identities and the effects of
topology in the patterns of chiral symmetry breaking are given. The
numerical methods used in the simulations are discussed in section
\ref{sec_numerical_methods}. The numerical results are presented in
section \ref{sec_numerical_results} and the paper is concluded in 
section \ref{sec_conclusions}.

\section{Lattice formulation}
\label{sec_lattice_formulation}

In this section, the ${\cal N}=1$, SU(2) SYM lattice action and operators are
presented.  The approach is similar in spirit as to the case of
Wilson fermions
\cite{Curci_Veneziano,Montvay,Donini}. The DWF formulation for this
theory is identical to \cite{Neuberger_fermions} and
\cite{Kaplan_Schmaltz}. It is presented below for the convenience of
the reader and in order to establish notation.

The ${\cal N}=1$, SU(2) SYM theory is an SU(2) gauge theory with Majorana
fermions in the adjoint representation. As such, the fermionic path
integral results in the analytic square root of the corresponding
Dirac determinant. This then is the Pfaffian of an antisymmetric
matrix that has the same determinant as the Dirac operator. On the
lattice, the Dirac operator can be defined using Wilson's approach as
in \cite{Curci_Veneziano,Montvay,Donini} or the DWF approach as in
\cite{Neuberger_fermions} and
\cite{Kaplan_Schmaltz}.

The partition function is:
\begin{equation}
Z = \int [dU] \int [d\Psi] \int [d\Phi] e^{-S}
\label{Z}
\end{equation}
$U_\mu(x)$, $\mu=1,2,3,4$ is the four-dimensional gauge field in the
fundamental representation, $\Psi(x,s)$ is a (real) five-dimensional
Majorana field in the adjoint representation and $\Phi(x,s)$ is a
(real) five-dimensional bosonic Pauli Villars (PV) type field with the
same indices as the Majorana field. $x$ is the coordinate in the
four-dimensional space-time box with extent $L$ along each of the four
directions. The boundary conditions along these directions are taken
to be periodic for all fields.  The coordinate of the fifth direction
is $s=0,1, \dots, L_s-1$, where $L_s$ is the size of that direction
and is taken to be an even number.  The action $S$ is given by:
\begin{equation}
S = S_G(U) + S_F(\Psi, U) +
S_{PV}(\Phi, U) \ .
\label{action}
\end{equation}

$S_G(U)$ is the pure gauge part and is defined
using the standard single plaquette action of Wilson:
\begin{equation}
S_G = \beta \sum_p ( 1 - {1 \over 2} {\rm Re Tr}[U_p])
\label{action_G}
\end{equation}
where $\beta = 4/g^2$ and $g$ is the gauge coupling.

The fermion part $S_F(\Psi, U)$ is given by:
\begin{equation}
S_F = - \sum_{x,x^\prime,s,s^\prime} \Psibar(x,s) D_F(x,s; x^\prime,
s^\prime) \Psi(x^\prime,s^\prime)
\label{action_F}
\end{equation}
where $D_F$ is the DWF Dirac operator in the form of \cite{Furman_Shamir}.
Specifically it is:
\begin{equation}
D_F(x,s; x^\prime, s^\prime) = \delta(s-s^\prime) D\slash(x,x^\prime)
+ D\slash^\bot(s,s^\prime) \delta(x-x^\prime)
\label{D_F}
\end{equation}
\begin{eqnarray}
D\slash(x,x^\prime) &=& {1\over 2} \sum_{\mu=1}^4 \left[ (1+\gamma_\mu)
V_\mu(x) \delta(x+\hat\mu - x^\prime) + (1-\gamma_\mu)
V^\dagger_\mu(x^\prime) \delta(x^\prime+\hat\mu - x) \right] \nonumber \\
&+& (m_0 - 4)\delta(x-x^\prime)
\label{Dslash_F}
\end{eqnarray}
\begin{equation}
D\slash^\bot(s,s^\prime) = \left\{ \begin{array}{ll} 
P_R \delta(1-s^\prime) - m_f P_L \delta(L_s-1 - s^\prime) - \delta(0 - s^\prime) & s=0 \\ 
P_R \delta(s+1 - s^\prime) + P_L \delta(s-1 - s^\prime) - \delta(s-s^\prime) & 0 < s < L_s-1 \\ 
-m_f P_R \delta(0-s^\prime) + P_L \delta(L_s-2 - s^\prime) - \delta(L_s-1 - s^\prime) & s = L_s -1
\end{array}
\right. 
\label{Dslash_perp_f}
\end{equation}
\begin{equation}
P_{R,L} = { 1 \pm \gamma_5 \over 2}
\end{equation}
where $V$ is the gauge field in the adjoint representation.
It is related to the field in the fundamental representation by 
(see for example \cite{Montvay}):
\begin{equation}
[V_\mu(x)]_{a,b} = 2 {\rm Tr} [U^\dagger_\mu(x) T^a U_\mu(x) T^b]
\label{adj_fnd}
\end{equation}
and
\begin{equation}
V_\mu(x) = V^{*}_\mu(x) = [V^{-1}_\mu(x)]^T
\label{adj_properties}
\end{equation}
where $T^a = {1 \over 2} \sigma^a$ with $\sigma^a$ the Pauli
matrices.
In the above equations $m_0$ is a five-dimensional mass representing
the ``height'' of the domain wall and it controls the number of
light flavors in the theory. In order to get one light species
in the free theory one must set $0<m_0<2$
\cite{Kaplan}. The parameter $m_f$ explicitly mixes the two
chiralities and as a result it controls the bare fermion mass of the
four-dimensional effective theory. The dependence of the bare fermion
mass on $m_0$ and $L_s$ is discussed in section \ref{sec_dwf_params}.

The fermion field $\Psibar$ is not independent but is related to
$\Psi$ by the equivalent of the Majorana condition for this
5-dimensional theory \cite{Kaplan_Schmaltz}:
\begin{equation}
\Psibar = \Psi^T C R_5
\label{5D_Majorana_cond}
\end{equation}
where $R_5$ is a reflection operator along the fifth direction and 
$C$ the charge conjugation operator in Eucledean space which
can be set to:
\begin{equation}
C = \gamma_0 \gamma_2 \ .
\label{charge_conj}
\end{equation}
Therefore, the fermion action can also be written as:
\begin{equation}
S_F = - \sum_{x,x^\prime,s,s^\prime} \Psi^T(x,s) 
M_F(x,s; x^\prime,s^\prime) 
\Psi(x^\prime,s^\prime)
\label{action_F1}
\end{equation}
where
\begin{equation}
M_F(x,s; x^\prime,s^\prime) = C R_5 D_F(x,s; x^\prime,s^\prime) 
\label{antisym_frm_matrix}
\end{equation}
is an antisymmetric matrix as can be easily checked \cite{Neuberger_fermions}.
As a result the fermionic integral gives the anticipated
Pfaffian:
\begin{equation}
\int [d\Psi] e^{-S_F} = {\rm Pf}(M_F)  \ .
\label{frm_pfaffian}
\end{equation}
Because $\det(C R_5) = 1$ one also has that $\det( M_F)  = \det( D_F ) $
and therefore:
\begin{equation}
{\rm Pf}(M_F) = \sqrt{\det( D_F )}  \ .
\label{frm_sqrt_det}
\end{equation}

The Pauli-Villars action $S_{PV}$ is designed to cancel the
contribution of the heavy fermions \cite{NN1}. Viewing the extra
dimension as an internal flavor space \cite{NN1} one can see that
there are $L_s-1$ heavy fermions with masses near the cutoff and one
light fermion. The PV subtraction subtracts the $L_s$ heavy particles.  As
was pointed in \cite{Neuberger_fermions} this amounts to a ``double''
regularization of the light degree of freedom, first by the lattice
and then by the PV field.  The form of the PV subtraction
used here is as in \cite{PMV_Schwinger} and is given by:
\begin{equation}
S_{PV} =
\sum_{x,x^\prime,s,s^\prime} \Phi^T(x,s) 
M_F[m_f=1](x,s; x^\prime, s^\prime) \Phi(x^\prime,s^\prime)  \ .
\label{action_PV}
\end{equation}
The integral over the PV fields results in:
\begin{equation}
\int [d\Phi] e^{-S_{PV}} = {1 \over {\rm Pf}(M_F[m_f=1])}  \ .
\label{frm_pfaffian_pv}
\end{equation}

Green functions in this work are measured using
four-dimensional fermion fields
constructed from five-dimensional fermion fields
using the projection prescription \cite{Furman_Shamir}:
\begin{eqnarray} 
\chi(x)    &=& P_R \Psi(x,0) + P_L \Psi(x, L_s-1) \nonumber \\
\chibar(x) &=& \Psibar(x,L_s-1) P_R + \Psibar(x, 0) P_L  \ .
\label{projection}
\end{eqnarray} 
In the $L_s \to \infty$ limit of the theory these operators
directly correspond to insertions in the overlap of appropriate creation
and annihilation operators \cite{NN1}. 

Using eq. \ref{5D_Majorana_cond} and \ref{projection} the Majorana
condition on the four-dimensional fermion field is:
\begin{equation}
\chibar = \chi^T C  \ .
\label{4D_Majorana_cond}
\end{equation}
Because this is the correct condition for a four-dimensional field
one can see that the definition in eq. \ref{5D_Majorana_cond}
not only produces an antisymmetric fermion matrix $M_F$ but is
also consistent with the projection prescription in eq. \ref{projection}
as expected.

\section{Analytical Considerations}
\label{sec_analy}

In this section some analytical considerations are presented.  In the
${\cal N}=1$ SYM theory, a gluino mass term is the only relevant operator that
can break supersymmetry and is also the only relevant operator that
can break (at the classical level) the $U(1)_A$ symmetry. Therefore,
the two symmetries are intimately related to the mechanisms that can
introduce a bare gluino mass term. These mechanisms depend on the
``extra'' regulator parameters $m_0$ and $L_s$.  This is discussed
below. Next the fate of the $U(1)_A$ chiral symmetry and the effects of
topology are presented.  The chiral and
supersymmetric Ward identities are derived in the last subsection.

\subsection{The ``extra'' DWF parameters}
\label{sec_dwf_params}

DWF introduce two extra parameters, the size of the fifth direction $L_s$ and
the domain wall height or five-dimensional mass $m_0$. These two parameters
together with the explicit mass $m_f$ control the bare fermion mass $\meff$.
In the free theory one finds \cite{PMV_Schwinger}:
\begin{equation}
\meff = m_0 (2 - m_0) \left[ m_f + (1-m_0)^{L_s}\right], \ \ \ \ 0 < m_0 < 2  \ .
\label{meff_free}
\end{equation}
In the interacting theory one would expect that $m_0$ as well as its
range of values will be renormalized. From the above equation one
can see that for the free theory the value of $m_0=1$ is optimal in
the sense that finite $L_s$ effects do not contribute to $\meff$.
In the interacting
theory one would expect that there is no such ``optimal value'' since,
in a heuristic sense, $m_0$ will fluctuate. 
For a more detailed analysis please see \cite{Shamir_weak_coupling}.
Then one would like $L_s$ 
to be large enough so that the second term in eq. \ref{meff_free} will 
be small allowing for simulations at reasonably small masses and/or
for dependable extrapolations to the $m_f \to 0$, $L_s \to \infty$ limit.

The effects of finite $L_s$ on the chiral symmetry can be best
understood in the overlap formalism \cite{NN1}.  In that formalism a
transfer matrix $T$ along the extra direction is constructed. Because
the gauge fields are not changing along that direction the product of
transfer matrices simply results in $T^{L_s}$. For $L_s = \infty$ this
is a projection operator that projects the reference vacuum state to a
ground state. The fermion determinant is then the overlap of the
reference vacuum state with that ground state. In \cite{NN1} it was
shown that, as a lattice gauge field configuration changes, from say
the zero topological sector to sector one, an eigenvalue 
(or a degenerate set of eigenvalues)
of the corresponding Hamiltonian $H$ changes sign. 
As a result, the filling
level of the ground state becomes different from that of the reference
vacuum state. Then the overlap is zero indicating the presence of an
exact zero mode. This remarkable property is maintained to a good
degree even at finite $L_s$ as was found in
\cite{DWF_zero_modes_Columbia}. Unfortunately, this property is also
the reason for most of the difficulties with DWF.  As the eigenvalue
of the Hamiltonian $H$ changes sign it crosses zero.  In such a
configuration the transfer matrix has an eigenvalue equal to one and
therefore even at $L_s=\infty$ there is no decay along the extra
direction, the two chiralities do not decouple, and chiral symmetry
can not be restored. Fortunately, configurations for which $H$ has an
exact zero eigenvalue (for a given $m_0$)
are of measure zero \cite{NN1,Furman_Shamir} and
therefore are of no consequence.  However, configurations in their
neighborhood are not of measure zero and such configurations will
exhibit very slow decay rates. Therefore, in order to restore chiral
symmetry, very large values of $L_s$ may be needed. Since one would
expect that the neighborhoods of such configurations are suppressed
closer to the continuum limit this problem should become less severe
as that limit is taken.  This has been observed in numerical
simulations of the Schwinger model \cite{PMV_Schwinger}, of full QCD
\cite{DWF_dyn_QCD_Columbia}, and of quenched QCD
\cite{DWF_quenched_QCD_Columbia,DWF_quenched_QCD_RBC,DWF_quenched_QCD_Japan,DWF_quenched_QCD_Jlab}.

In the region where it makes sense to parameterize these effects
by a residual mass in an effective action it has been found that:
\begin{equation}
\meff = c_0  m_f + \mres, \ \ \ \  \mres = c_1  \exp(-c_2  L_s)
\label{meff_int}
\end{equation}
where for dynamical QCD at the currently accessible lattice spacings
the decay is found to be $c_2 \approx
0.02$ \cite{DWF_dyn_QCD_Columbia}. For quenched QCD the situation is better
because current computing resources can simulate lattices with smaller
lattice spacing. There, a value of $c_2 \approx 0.1$ is found
\cite{DWF_quenched_QCD_Columbia,DWF_quenched_QCD_RBC,DWF_quenched_QCD_Japan}. 
Also in these studies the value of $c_2$ was
a weakly changing function of $m_0$ indicating that for practical
purposes there is no optimal value of $m_0$.

In the case of the ${\cal N}=1$ SYM SU(2) theory the Hamiltonian corresponding
to the five dimensional transfer matrix has eigenvalues that are
doubly degenerate because the fermion fields are in the adjoint
representation \cite{EHN_adjoint_index}.  Therefore when there is a
``topology'' change two eigenvalues will have to cross through zero
(as compared to one for fundamental fermions).  This may make this
theory harder to study than QCD in the sense that larger $L_s$ values
may be required. On the other hand, since no massless Goldstone
particles are expected, the sensitivity of the spectrum on $L_s$ may be
considerably milder. In any case, in this paper the only fermionic
observable that will be discussed is the gluino condensate. This
quantity is known to approach its $L_s \to \infty$ limit with faster
decay rates than the ones in $\mres$ (for a discussion and results for
full QCD see \cite{DWF_dyn_QCD_Columbia}; there the decay rate for the chiral
condensate was about five times faster than that for $\mres$).

As was discussed above, the range of $m_0$ is renormalized by the
interactions. It has been found that as the lattice spacing increases
and one moves away from the continuum limit this range shrinks in size
and for currently accessible spacings in QCD that range is about 
$[1.4, 2.0]$. As one moves even farther away from the
continuum limit this range can shrink to zero and then it will not be
possible to have light DWF modes 
\cite{Neuberger_fermions,Brower_Svetitsky}. However, it must be 
emphasized that for as long as the range of allowed values of $m_0$ is 
not of zero size the overlap formalism, 
although it does not specify 
how it is approached,
guarantees the existence of the $L_s \to \infty$ limit.
In this work, $m_0=1.9$ and, as it will be shown in section 5,
the behavior of the gluino condensate vs. $L_s$ is consistent
with an exponential ansatz.

\subsection{Chiral symmetry and topology}
\label{sec_topology}

Fermions in the adjoint representation of the SU(N) gauge group
have a Dirac operator with index:
\begin{equation}
2 N \nu
\label{index}
\end{equation}
where $\nu$ is the winding of the background field configuration. 
Classical instantons have integer winding and
they cause condensation of operators with $2 N$ Majorana
fermions. This results in the breaking of the $U(1)_A$ chiral symmetry
down to the $Z_{2N}$ symmetry by the corresponding anomaly. The
remaining $Z_{2N}$ symmetry may break spontaneously down to
$Z_2$ \cite{SYM_gluino_cond}. Mechanisms for this further breaking
have been explored for example in 
\cite{Cohen_Gomez,Schaefer,Mattis_etal}
where instantons and fractionally charged objects 
such as torons \cite{tHooft_torons} or caloron monopole constituents
\cite{Kraan_VanBaal} were investigated as the source of 
this symmetry breaking.

Since in a toroidal geometry fractional winding numbers are 
possible \cite{tHooft_torons},
the partition function of the full theory can be expressed as:
\begin{equation}
Z(\theta) = \sum_\nu e^{i \nu \theta} Z_\nu, \ \ \ \ \ \nu = 0, \pm 1/N, \pm 2/N, \cdots
\label{partition_function}
\end{equation}
where $\theta$ is the vacuum angle and $Z_\nu$ is the partition
function on the sector with winding $\nu$. For the theory with a soft
breaking by a mass $m_f$
the interplay of the volume and mass in the
formation of the gluino condensate has been analyzed in
\cite{Leutwyler_Smilga}.  The reader is referred to that reference for
a very nice presentation on the subject.  Assuming a mass gap is
present in the theory the authors of \cite{Leutwyler_Smilga} show
that non-zero contributions to the 
gluino condensate $\chibarchi$ come almost
exclusively from the $\nu= 1/N$ sector if $m_f \times V \times \chibarchi \ll 1$.
On the other hand, if $m_f \times V \times \chibarchi \gg 1$
all sectors contribute to a non-zero condensate.

The above considerations result in an unusual picture.  If the
infinite volume limit is taken (followed by the massless limit) it is
possible that a gluino condensate will form due to spontaneous
breaking of the discrete symmetry $Z_{2N}$ down to $Z_2$. On the other
hand, at a finite volume and zero mass a gluino condensate can form due
to the presence of fractional winding configurations. Since the volume is
finite, this can not be the result of spontaneous symmetry
breaking. Instead, it is similar to symmetry breaking due to
topological effects as, for example, in one flavor QCD.  As pointed
above the size of $m_f \times V \times \chibarchi $ controls which ``scenario'' takes
place.

On the lattice there is no clear definition of topology.  The path
integral over the SU(N) group space generates configurations of all
possible windings. In order for the lattice
theory to be able to reproduce phenomena that relate to topology it is
essential that the lattice Dirac operator obeys the index theorem in a
statistical sense. This is highly non--trivial since it is obviously
related to the doubling problem. Traditional fermions (Wilson or
staggered) do not exhibit exact zero modes at finite lattice spacing.  
On the other hand, as mentioned in section
\ref{sec_dwf_params}, DWF at $L_s = \infty$ have exact zero modes and at
finite $L_s$ have robust zero modes to a good approximation 
\cite{DWF_zero_modes_Columbia}. An
approximate form of the index theorem has been found to be obeyed for
fundamental fermions in the overlap formulation in quenched SU(2)
\cite{Narayanan_Vranas} and in quenched SU(3)
\cite{EHN_su3_index}.

The index of adjoint fermions in the overlap formulation in quenched
SU(2) has been studied in \cite{EHN_adjoint_index}. In that work it
was pointed out that the overlap Dirac operator for adjoint fermions
in the SU(2) gauge group is necessarily even--valued.  Then the
question posed by the authors of \cite{EHN_adjoint_index} was whether
or not all even values are realized or only values that are multiples
of four are present.  The latter case corresponds to configurations
with instantons.  The former case corresponds to fractional winding
numbers.  Configurations with fractional winding were found and their
presence persisted as the lattice spacing was decreased

In this paper DWF are used at finite $L_s$ and therefore some of the
clarity present in the $L_s=\infty$ case will be lost. However, the
full theory (including the fermion determinant) is studied here. 
Furthermore, it is
interesting to see if at a small volume and zero mass the gluino
condensate still forms and if it does to what extent its value is due
to zero mode effects. The numerical results
are presented in section \ref{sec_numerical_results}.

\subsection{Ward identities}
\label{sec_ward_identities}

As discussed in the introduction and in section
\ref{sec_analy}, the DWF formulation of the ${\cal N}=1$ SU(2) SYM theory
at the $L_s \to \infty$
limit is expected to preserve the $U(1)_A$ chiral symmetry (at the
classical level) and break supersymmetry only by irrelevant operators.
Since the DWF formulation contains many more fields than the continuum
theory, one may naturally wonder what are the SUSY transformations in
terms of these fields. In particular, while the continuum theory has a
single Majorana fermion the DWF lattice theory contains $L_s$ Majorana
fermions and $L_s$ corresponding PV fields.  Since all these fields,
except for one Majorana fermion, have masses near the cutoff, one can
expect that the SUSY transformations should only transform the gauge
field and the light Majorana fermion represented by the boundary field
$\chi$ of eq. \ref{projection}. Similarly, the chiral symmetry
transformations should only involve the field $\chi$. However, one
should expect that this choice of SUSY and chiral transformations is
not unique. For example, see \cite{Furman_Shamir} for a different
choice of QCD chiral transformations that involve all fermion fields
in one half of the fifth direction transforming vectorially and all
fermions in the other half also transforming vectorially but with
opposite charge.  That choice could also be appropriate here for the
chiral transformations, but it may make the SUSY ones more
complicated.

As a first step in deriving the Ward identities, the fermionic part of
the action in eq. \ref{action_F} is rewritten in terms of the boundary
field $\chi$:
\begin{equation}
S_F = S_{F_0} + S_{F_\chi}
\label{sf0_sfchi}
\end{equation}
where $S_{F_0}$ does not depend on the field $\chi$ and
\begin{eqnarray}
S_{F_\chi} = - \sum_{x,x^\prime}
             && \!\!\!\!\!\! 
                  \left[ \  \chibar(x) D\slash_N(x,x^\prime) \chi(x^\prime)
                - m_f \chibar(x) \delta(x,x^\prime) \chi(x^\prime) \right.\nonumber \\
             && \!\!\!\!\!\! 
                - \chibar(x) B(x,x^\prime) \phi(x^\prime) 
                - \left. \phibar(x) B(x,x^\prime) \chi(x^\prime) \right]
\label{sfchi}
\end{eqnarray}
where
\begin{eqnarray} 
\phi(x)    &=& P_R \Psi(x, L_s-1) + P_L \Psi(x, 0) \nonumber \\
\phibar(x)  &=& \Psibar(x, 0) P_R + \Psibar(x, L_s-1) P_L
\label{wrong_projection}
\end{eqnarray} 
are the ``wrong'' projected fields in the sense that they are defined
on the opposite wall from where the corresponding light mode is
localized. If indeed there is localization one
would expect that in the $L_s \to \infty$ limit these fields
will have no overlap with the light mode. 
The operator $D\slash_N$ is the naive part of the four-dimensional 
Wilson operator in eq. \ref{Dslash_F} and $B$ is the symmetry breaking part
($B$ is the equivalent of $B$ in \cite{NN1,Furman_Shamir}):
\begin{equation}
D\slash_N(x,x^\prime) = 
{1 \over 2} \sum_{\mu=1}^4 \gamma_\mu \left[ V_\mu(x) \delta(x+\hat\mu - x^\prime)
- V^\dagger_\mu(x^\prime) \delta(x^\prime+\hat\mu - x) \right]                  
\label{Dslash_F_N}
\end{equation}
\begin{equation}
B(x,x^\prime) = 
(5 - m_0) \delta(x - x\prime)
- {1 \over 2} \sum_{\mu=1}^4 \left[ V_\mu(x) \delta(x+\hat\mu - x^\prime)  
+ V^\dagger_\mu(x^\prime) \delta(x^\prime+\hat\mu - x) \right]
\ .
\label{D_F_B}
\end{equation}
These operators have the following properties:
\begin{eqnarray} 
\{ D\slash_N, \gamma_5 \} &=& 0, \ \ \ \ \  D\slash_N^\dagger = - D\slash_N \nonumber \\
\left[ B, \gamma_5 \right] &=& 0, \ \ \ \ \  B^\dagger = B^T = B 
\label{Dslash_F_N_I_properties}
\end{eqnarray} 

First the Ward identity corresponding to the $U(1)_A$ symmetry is derived.
The symmetry transformations are:
\begin{eqnarray} 
\delta_A \chi(x)     &=& i \alpha(x) \gamma_5   \chi(x)   \nonumber \\
\delta_A \chibar(x)  &=& i \alpha(x) \chibar(x) \gamma_5 
\label{chiral_transform}
\end{eqnarray} 
where $\alpha(x)$ is an infinitesimal real number and $\delta_A$
symbolizes the change under the chiral transformation. Then the Ward identity is:
\begin{equation}
\langle \Delta_\mu  J_\mu(x) {\cal O}(y) \rangle = 2 m_f \langle J_5(x) {\cal O}(y) \rangle
+ 2 \langle J_B(x) {\cal O}(y) \rangle + i \langle \delta_A {\cal O}(y) \rangle
\label{chiral_WI}
\end{equation}
where the backward difference is defined as $\Delta_\mu f(x) \equiv  f(x) - f(x-\mu)$. 
The currents are:
\begin{equation}
J_\mu(x) = \chibar(x) \gamma_5 \gamma_\mu V_\mu(x) \chi(x+\mu)
\label{j_mu}
\end{equation}
\begin{equation}
J_5(x) = \chibar(x) \gamma_5 \chi(x)
\label{j_5}
\end{equation}
\begin{equation}
J_B(x) = \sum_y \chibar(x) \gamma_5 B(x,y) \phi(y)  \ .
\label{J_B}
\end{equation}

If in the above Ward identity ${\cal O}(y) = J_5(y)$ one gets
\begin{equation}
\Delta_\mu \langle J_\mu(x) J_5(y) \rangle = 2 m_f \langle J_5(x) J_5(y) \rangle
+ 2 \langle J_B(x) J_5(y) \rangle -2  \langle \chibar(y)\chi(y) \rangle
\label{chiral_WI_J5}
\end{equation}
In this identity the term with $J_B$ will be responsible for
producing the ABJ anomaly in the $L_s \to \infty$ limit.
On the other hand, if $L_s$ is kept finite this term is similar to the one
for Wilson fermions which, besides producing the ABJ anomaly,
also produces a mass redefinition. For an analysis
of QCD with DWF at finite $L_s$ see \cite{DWF_quenched_QCD_RBC}.

As mentioned earlier these chiral transformations are different than the ones
in \cite{Furman_Shamir}. If the transformations relevant for a non-singlet 
current in QCD were done on the fields $\chibar$, $\chi$, one obtains
a Ward identity exactly as in \cite{Furman_Shamir} but with the
currents $A^a_\mu(x)$ and $J^a_{5q}(x)$ replaced with:
\begin{equation}
A^a_\mu(x) = {1 \over 2} 
\left[\chibar(x)     \gamma_5 \gamma_\mu \lambda^a U_\mu(x)         \chi(x + \mu)
    + \chibar(x+\mu) \gamma_5 \gamma_\mu \lambda^a U^\dagger_\mu(x) \chi(x) \right]
\label{A_mu}
\end{equation}
\begin{equation}
J^a_{5q}(x) =  {1 \over 2} \sum_y 
\left[\chibar(x) \gamma_5 \lambda^a B(x,y) \phi(y)
   +  \phibar(y) \gamma_5 \lambda^a B(y,x) \chi(x) \right]  \ .
\label{J_5q}
\end{equation}

The derivation of the SUSY Ward identity is similar to the one for
Wilson fermions. One can use the existing calculations for Wilson 
fermions \cite{Curci_Veneziano,Donini,Taniguchi}
to elucidate the differences between the two formalisms.  
Here the derivation in \cite{Taniguchi} will be followed. The symmetry
transformations are as in \cite{Taniguchi} and commute with parity.

The change of the pure gauge action with respect to the transformation of the
gauge field is of course the same.  In
terms of the symmetry breaking part of the Ward identity it
contributes a term denoted below by $X_2(x) + X_3(x)$ 
where $X_2$, $X_3$ are as in \cite{Taniguchi}.
This term breaks SUSY because of the explicit
breaking of the Lorentz symmetry. Using improved pure gauge lattice
actions can alleviate the effects of this breaking.  Such an
improvement is not considered here.

The change of the fermion and Pauli-Villars parts of the DWF action 
with respect to the transformation of the gauge fields produces terms 
for all $L_s$ slices. In particular the variation of the fermion
matrix $D_F$ of eq. \ref{D_F} with respect to the gauge field is:
\begin{equation}
\delta_U D_F(x,x^\prime; s,s\prime)(m_f) = \delta(s-s^\prime) \delta_U D\slash(x,x^\prime)
\label{delta_D_F}
\end{equation}
One sees that $\delta D_F$ is independent of $m_f$ and is diagonal in the
fifth direction. Furthermore $\delta D\slash(x,x^\prime)$ is the same as the variation
of the Wilson operator. Therefore, this variation contributes to the symmetry breaking part of
the Ward identity the terms:
\begin{equation}
X_4^F(x) = \sum_s X_4^F(x,s)
\label{X_4_F_DWF}
\end{equation}
and
\begin{equation}
X_4^{PV}(x) = \sum_s X_4^{PV}(x,s)
\label{X_4_PV_DWF}
\end{equation}
where $X_4^F(x,s)$ is as $X_4$ in \cite{Taniguchi}
except that the four-dimensional Wilson fermion 
fields that have their spin indices contracted are replaced by
$\Psibar(x,s)$, $\Psi(x,s)$
while the other Wilson fermion field is replaced by
$\chi(x)$. Similarly $X_4^{PV}(x,s)$ is as $X_4$ in \cite{Taniguchi}
except that the
four-dimensional Wilson fermion fields that have their spin indices
contracted are replaced by the Pauli-Villars fields $\Phi^T(x,s) C R_5$, $\Phi(x,s)$,
the other Wilson fermion field is replaced by $\chi(x)$ and the sign of the second term
in $X_4$ is minus instead of plus due to the commutativity of the Pauli-Villars fields.

The change of the action with respect to the fermion field
transformations can be partially
deduced from the corresponding Wilson fermion calculation. Since this
transformation only involves the action $S_{F_\chi}$ in
eq. \ref{sfchi}, one can observe that the first two terms of that
action are identical with the action of naive fermions (Wilson
fermions with $r=0$). These will contribute identical terms as the
$r=0$ part of the Wilson action. They contribute to the divergence of
the SUSY current and to the mass term of the Ward identity given
below.
Finally, the transformation of the last term of the action $S_{F_\chi}$
in eq. \ref{sfchi} is easy to calculate and is denoted by $X_1(x)$:
\begin{equation}
X_1(x) = - P^a_{\rho \sigma}(x) \sigma_{\rho \sigma} B^{ab}(x,x^\prime) \phi^b(x^\prime)  \ .
\label{X_1}
\end{equation}
This term is closely related to $X_1$ of \cite{Taniguchi}. 

The Ward identity is:
\begin{equation}
\langle \Delta_\mu  S_\mu(x) {\cal O}(y) \rangle = 
m_f \langle D_s(x) {\cal O}(y) \rangle
+ \langle X_S(x) {\cal O}(y) \rangle - \langle \delta_S {\cal O}(y) \rangle
\label{susy_WI}
\end{equation}
where the supersymmetric current $S_\mu$ and the quantity $D_S$ are as in \cite{Taniguchi}.
The symmetry breaking term $X_S(x)$ is also similar to the one in \cite{Taniguchi}:
\begin{equation}
X_S(x) =  X_1(x) + X_2(x) + X_3(x) + X_4^F(x) - X_4^{PV}(x)
\label{X_S}
\end{equation}

As mentioned above the symmetry breaking term $X_2(x) + X_3(x)$ is due
to the breaking of Lorentz symmetry by the lattice. The
$X_4^F(x)$ and $X_4^{PV}(x)$ terms break the symmetry as in
Wilson fermions. These terms do not cancel each other exactly 
\footnote{We thank Y. Shamir for pointing this out to us}.  
However, one would
expect large cancellations of heavy modes. The terms in $X_4^F(x,s)$
that are proportional to the Wilson parameter involve fields that
couple to the light modes by an amount that is exponentially small in
$L_s$.  One would expect these terms to be nearly canceled by the
corresponding Pauli-Villars terms resulting in exponentially small
contributions.  The remaining terms that involve fields away from the
relevant domain walls should also yield similar cancellations. As a
result the only terms that should make significant contributions
should be the ones that involve fields of the ``correct'' chirality
near the domain walls. These
few terms would couple to the light modes and be further regularized
by the corresponding Pauli-Villars terms. Clearly this analysis
of cancellations is heuristic. A detailed calculation using for example
perturbation theory or transfer matrix methods would
be interesting but it is beyond the 
scope of this paper.

Finally, the symmetry breaking term $X_1(x)$ involves the field
$\phi(x)$ that is expected to have no overlap with the light mode in
the $L_s \to \infty$ limit.  If $L_s$ is finite then DWF are similar
to Wilson fermions and an analysis as in \cite{Curci_Veneziano} should
indicate that this term is responsible for the same mass redefinition
as the one in the chiral Ward identity.

\section{Numerical methods}
\label{sec_numerical_methods}

As can be seen from section \ref{sec_lattice_formulation} 
the ${\cal N}=1$ SU(2) SYM theory
can be simulated as a theory with 0.5 flavors of Dirac fermions in the
adjoint representation. An efficient and popular algorithm that can be
used to simulate any number of flavors is the hybrid molecular dynamics R
(HMDR) algorithm of
\cite{Gottlieb}. Because of the Grassmann nature of fermions
these algorithms need to invert the matrix $D_F$ of eq. \ref{D_F}.
That matrix is not Hermitian. This is a problem since some of the more efficient
inversion algorithms require the matrix to be Hermitian. However, because:
\begin{equation}
\gamma_5 R D_F R \gamma_5 = D_F^\dagger
\label{g5_R_D_F}
\end{equation}
one has that:
\begin{equation}
\det[D_F]^2 = \det[D_F D_F^\dagger] \ .
\label{det_D_F_sq}
\end{equation}
Then one can invert the
Hermitian matrix $D_F D_F^\dagger$ and then use the HMDR algorithm to take
the appropriate power so that the desired number of flavors is
simulated.  This method is adopted here and the 0.25 power is taken in
order to go from a theory with two Dirac fermions to a theory with one
Majorana fermion.  In other words, the fermion determinant that is used
in the simulation is:
\begin{equation}
(\det[D_F D_F^\dagger])^{0.25} = |\det[D_F]|^{0.5} = \det[D_F]^{0.5}
\label{R_det}
\end{equation}
where in the last equality use was made of the fact that for
non-negative $m_f$ $\det[D_F]$ is also non-negative \cite{Neuberger_fermions}.
This approach was also taken in \cite{Donini} for Wilson fermions.
For an approach that uses Wilson fermions and the multibosonic
algorithm \cite{multi_boson} instead of the HMDR algorithm see
\cite{Montvay}. However, as mentioned earlier in the case of Wilson 
fermions the last equality in eq. \ref{R_det} is not true for all gauge 
field configurations.  

The HMDR algorithm uses molecular dynamics methods in order to produce
the correct statistical ensembles. Because the molecular dynamics step
size $\delta\tau$ is finite discretization errors are introduced.
There are two ways one can deal
with this problem. One is to simulate at various values of
$\delta\tau$ and then extrapolate to $\delta\tau=0$.  Another method
is to use $\delta\tau$ small enough so that the errors are negligible
when compared with the statistical errors.

In order to ensure this, one can simulate the two Dirac flavor theory
at the same parameters and same $\delta\tau$. For the two flavor
theory, one has a local action and therefore, at the end of the
evolution, one can employ a Metropolis accept-reject step.  Then the
finite $\delta\tau$ errors are ``converted'' to a non-ideal acceptance
rate and in effect they are reflected in the final statistical errors.
This is the exact hybrid Monte Carlo $\Phi$ (HMC$\Phi$) algorithm of
\cite{Duane,Gottlieb}.  Therefore the acceptance rate is an indication
of the size of the finite $\delta\tau$ errors in the HMD integration.
By simulating the two Dirac flavor theory with (HMC$\Phi$) one can set
$\delta\tau$ so that the acceptance rate is high, say $\approx 90\%$.
Since the coefficient of the finite $\delta\tau$ errors is
proportional to the number of flavors one would expect that for
0.25 flavors the errors would be small and at the few percent level.

The only fermion observable measured in this work is the gluino
condensate. By inserting appropriate source terms as in \cite{Donini}
the gluino condensate was measured as the trace of $D_F^{-1}$ with
spin and fifth-direction indices restricted as dictated by
eq. \ref{projection} The trace was calculated using a standard
stochastic method.  All inversions in this work were done using the
conjugate gradient (CG) algorithm. An even-odd preconditioned form of
the matrix $D^\dagger_F D_F$ was inverted. For more details on the numerical
algorithms and methods employed to DWF simulations see
\cite{PMV_Schwinger,DWF_dyn_QCD_Columbia}.

\section{Numerical results}
\label{sec_numerical_results}

\subsection{Simulation parameters}
\label{sec_simulation_params}

In all simulations the domain wall height was chosen to be $m_0=1.9$.
As mentioned in the previous section, the finite $\delta\tau$ errors
were kept to the few percent level by using a small $\delta\tau$.  
For all simulations the step size was set to
$\delta\tau=0.01$ and the trajectory length to $\tau=0.5$.
In order to confirm that this choice introduces finite step size errors
that are small compared to the statistical errors
an HMC$\Phi$ simulation for two Dirac flavors was run for
$L_s = 12$ and $m_f=0.04$. It produced an acceptance rate of $\approx 90\%$
suggesting that the finite $\delta\tau$ errors of the $0.5$ flavor theory 
are small.
Furthermore, an HMDR simulation was also run for two Dirac flavors
using the exact same parameters. The value of the gluino condensate 
obtained from these two simulations was the same within statistical errors.

The CG stopping condition for all simulations was set to $10^{-6}$ for
the evolution and to $10^{-8}$ for the calculation of
$\chibarchim$. The number of CG iterations varied between $\approx 100$
for $m_f=0.08$, $L_s=12$ and $250$ for $m_f=0.0$, $L_s=24$. 

The $8^4$ volume simulations were done with $\beta=2.3$.
The value of $\beta$ was chosen so
that one is not close to the point where the box size becomes too
small and a thermal transition takes place, but also not too deep in
the strong coupling regime where the finite $L_s$ effects become
severe. The transition point of the $N_t=8$ quenched theory 
is at $\beta = 2.5115(40)$ \cite{Fingberg}. In figure
1 the magnitude of the fundamental Wilson line $\langle |W| \rangle$ measured in
quenched simulations in an $8^4$ volume is plotted vs.  $\beta$. In
the quenched theory this is an order parameter.  As can be seen from
that figure, a rapid crossover takes place around $\beta=2.5$. In the
same figure the value of $\langle |W| \rangle$ from a simulation of the dynamical
theory at $\beta=2.3$ is also shown (cross).  The quenched and
dynamical values are very similar indicating that at $\beta=2.3$ the
dynamical theory is in a phase that ``confines'' fundamental sources.
Therefore, the box size is large enough to avoid finite temperature
effects that would of course spoil SUSY. Using the quenched theory as
a guide the $4^4$ simulations were done at $\beta=2.1$ since the
quenched transition at $N_t=4$ is known to take place 
at $\beta = 2.2986(6)$ \cite{Fingberg}. At
$\beta=2.1$ the lattice spacing is larger than at $\beta=2.3$.
However, the lattice sizes are small and do not allow
a reliable measurement of the lattice spacing. 
According to \cite{Creutz_SU2}, $\beta=2.1 - 2.3$ is in the
beginning of the weak coupling regime. Then if one uses the weak
coupling form in \cite{Creutz_SU2} one finds that the lattice spacing
at $\beta=2.1$ is about a factor of two larger than the one at
$\beta=2.3$.

In order to estimate the necessary number of thermalization sweeps two
simulations were run on an $8^4$ lattice at $\beta=2.3$, $L_s=12$ and
$m_f=0.04$. The first simulation used an initial configuration with
all gauge links set to the identity (ordered) and the other used an
initially random configuration (dis-ordered).  The evolutions in
``computer time'' are shown in figure 2.  As can be seen, the two
ensembles converged after about $100$ sweeps.  This number of
thermalization sweeps was then used in all other simulations which
were started from an ordered initial configuration. The number of
measurements after thermalization for all simulations is about $200$
with measurements done in every trajectory.  The gluino condensate was
measured with a single ``hit'' stochastic estimator.

\subsection{The gluino condensate at the chiral limit}
\label{sec_gluino_meas}

In order to be able to extrapolate to the chiral limit, corresponding
to $L_s \to \infty$ and $m_f=0$, the mass $m_f$ and the size of the fifth
direction $L_s$ was varied.  The results of all simulations are given in tables
\ref{tab_8888_data} and \ref{tab_4444_data} in the appendix. 
Three different methods were used to analyze the data and calculate
the gluino condensate in the chiral limit.

I. For fixed $L_s$, the data for $m_f=0.08,  0.06, 0.04, 0.02$
were fit to a function:
\begin{equation}
b_0 + b_1 m_f  \ .
\label{linear}
\end{equation}
This functional form is valid provided $m_f$ is small enough.
Otherwise, higher order terms must also be included.
The data and fits are shown in figure 3 and the results
of the fits are given in table \ref{tab_linear_fits}.
Then the extrapolated values $b_0$ were fit
to a form:
\begin{equation}
c_0 + c_1 \exp(-c_2 L_s) \ .
\label{exponential}
\end{equation}
This functional form is approximate but it is expected to be valid
close enough to the continuum and has been found to be consistent in
simulations of the Schwinger model \cite{PMV_Schwinger} and of 
QCD even at relatively large lattice spacings (see for example
\cite{DWF_dyn_QCD_Columbia}). The data and fit is shown
in figure 4 and the results of the fit are given in table
\ref{tab_exp_fits}.

II. For fixed $m_f$ the data for $L_s=12,16,20,24$ were fit to the form
of eq. \ref{exponential}. The data and fits are shown in figure 5 and
the results of the fit are given in table \ref{tab_exp_fits}.  Then
the extrapolated values $c_0$ were fit to the form of
eq. \ref{linear}.  The data and fit is shown in figure 6 and the
results of the fit are in table \ref{tab_linear_fits}.

III. Additional simulations were done for $m_f=0$ and
$L_s=12,16,20,24$.  The data were fit to the form of
eq. \ref{exponential}. The data and fits are shown in figure 7
and the results of the fit are in table \ref{tab_exp_fits}.

The $m_f \to 0$ and $L_s \to \infty$ extrapolated values of the gluino
condensate for each one of the above three methods are summarized in
table \ref{tab_extrap}. As can be seen, all values are consistent
within the statistical errors. This suggests that the
systematic errors inherent to the limited statistics and to
fits onto functions that represent the data only for a limited range
are small. Furthermore, it suggests that the fitting functions used
are consistent (please see subsection \ref{sec_fine_print} for more
discussion on the validity of these fitting functions).

\subsection{The telltale signals of topology in numerical simulations}
\label{sec_numerical_topology}

In order to investigate the issues discussed in section
\ref{sec_topology} the gluino condensate was also calculated in a
smaller $4^4$ lattice volume at $\beta=2.1$. It was measured only for $m_f=0$ and
method III above was used to extrapolate to the $L_s \to \infty$
limit. The data and fit are shown in figure 8 and the fit results
are given in table \ref{tab_exp_fits}. The $8^4$ data from figure 7
are presented again in this figure to aid comparison.
The value has decreased indicating that scaling is violated. However,
without more simulations at other lattice spacings and volumes one can
not conclude much from this result. The $\beta=2.1$ coupling is in the strong 
coupling region and furthermore the $4^4$ lattice volume is rather small.

However, it is interesting to notice that the parameter 
$V \times \chibarchi_{L_s \to \infty} \approx 8.4$
(a factor of 12 coming from the normalization of $\chibarchi$
has been included).
Since $m_f=0$ the effective mass $\meff$
gets its value from finite $L_s$ effects. As $L_s$ is increased
$\meff$ becomes small.  
From analysis of $\meff$ in strong coupling QCD \cite{DWF_dyn_QCD_Columbia} one 
would roughly guess that $\meff < 0.1$. Then
$[\meff \times V \times \chibarchi_{L_s \to \infty}] < 1$. 
In that case, the analysis of
\cite{Leutwyler_Smilga} can be followed and one would expect the value
of the condensate in the $4^4$ lattice to be mostly supported by
configurations with total winding of $\pm 1/2$.  Indeed, this can be
seen from figure 9.  In that figure the evolutions in ``computer
time'' are shown.  The ``spikes'' in the evolution are apparent and
they become more pronounced and less frequent as $L_s$ is increased
(and in effect $\meff$ is decreased).  This is exactly how the effect
of zero modes for small 
$[\meff \times V \times \chibarchi_{L_s \to \infty}]$
would present itself in a
numerical simulation of the dynamical theory. As the fermion mass is
made smaller, $\chibarchi$ is expected to receive most of its value
from sectors with winding $\pm 1/2$. However, in these sectors the
fermion determinant is very small because of the zero mode. Since the
probability for the algorithm to generate a gauge field configuration
is proportional to the fermion determinant one would expect that these
sectors will be visited less and less frequently as the effective mass is
decreased. When these sectors are visited the value of $\chibarchim$
will be very large (spikes) in order to compensate for the infrequent
sampling. In this way the presence of the zero mode in the observable
``balances'' the presence of the zero mode in the determinant.  As the
mass is made smaller one would have to increase the size of the
statistical sample in order to include enough of these increasingly
``rare'' but very large fluctuations.  For similar results in the Schwinger model and
QCD see \cite{PMV_Schwinger,DWF_zero_modes_Columbia}.

A histogram of the values of $\chibarchim$ is presented in figure 10
(solid line).  For small $L_s$ the effective mass is larger and
$\chibarchim$ is distributed with a symmetric looking distribution
around the mean value. However, for $L_s=40$ the effective mass is
smaller and the distribution has a more pronounced ``tail'' towards
larger values.  In order to investigate this further numerical
simulations at exactly the same parameters, but without the fermion
determinant (quenched theory) were done. The histograms from these
simulations are shown in the same figure for
comparison (dotted lines). One can observe that the absence of the fermion
determinant had the effect of shifting the distributions to higher
values. This is expected since configurations with small eigenvalues
that produce larger values of $\chibarchim$ are not suppressed anymore
and are produced more frequently. Also, one can observe that the number
of configurations with $\chibarchim$ larger than $\approx 0.007$ that
appeared as spikes in figure 9 have now increased in number. These
observations lend support to the presence of small near-zero
eigenvalues. Furthermore, configurations with fractional 
topological charge have already been found in quenched SU(2) simulations
at similar couplings \cite{EHN_adjoint_index}. It would be very interesting
to calculate the index for the configurations of figure 9 using the methods of
\cite{EHN_adjoint_index} and see to what extent
there is a correlation between fractional topological charge 
and the observed spikes. This correlation should be exact 
at $L_s \to \infty$ but it will be obscured at finite $L_s$ by the presence
of non-zero $\meff$. This investigation is beyond the scope of this work.

Furthermore, it should also be noted that on the $8^4$ lattice there are no
visible spikes up to $L_s=24$. This can be seen in figure 11.
Presumably this is because the product 
$[\meff \times V \times \chibarchi_{L_s \to \infty}]$ is probably much larger
than in the $4^4$ lattice. Again, this statement is not exact since the
value of $\meff$ was not measured.

These results are consistent with the discussion in section
\ref{sec_topology}. However, since even with $m_f=0$,  an $L_s$ extrapolation 
is essentially an extrapolation from non zero masses these results are
not necessarily the results of a simulation at exactly $L_s=\infty$.
It is still possible that if such a simulation were done one could have
found that the gluino condensate is zero. This could happen since in a
finite volume and zero mass the effects of spontaneous symmetry
breaking are absent and the zero mode effects alluded to above may not
be sufficient to sustain a non zero vacuum expectation value.  This
type of simulation is possible and can be done using the overlap
formalism \cite{NN1} or exact Neuberger fermions
\cite{Neuberger_fermions}.  However, if one is to maintain exact
chiral symmetry these methods will demand large computing resources.

\subsection{The fine print}
\label{sec_fine_print}

Perhaps the largest uncertainty in the analysis presented in the
previous subsections has to do with the assumption of exponential decay
as in eq. \ref{exponential}. For small enough lattice spacings and
large enough $L_s$ this behavior is expected to be true.  
All data presented in this work were
well represented by this ansatz.
However, as with any
numerical investigation, one can never completely disprove all other
possibilities. While such an exercise over all possible functions
would clearly be fruitless there are some alternative forms that may
be reasonable to consider since they are based on analytical
considerations.

Far from the continuum limit, the approach to the chiral
limit may become power law \cite{EHN_su3_index} or even
completely disappear \cite{Neuberger_fermions,Brower_Svetitsky}. 
In order to explore the possibility of power law behavior the 
$m_f=0$ data for the $8^4$ and $4^4$ volumes were fit to the form:
\begin{equation}
d_0 + d_1 L_s^{d_2} \ .
\label{plaw}
\end{equation}
The results of the fit are given in table \ref{tab_plaw_fits} (the
fits are not presented in any of the figures).  As can be seen from
that table the $\cdof$ of these fits is significantly larger than the
one of the corresponding exponential fits to the same data.

Another possibility is decay to zero with two different exponential
decay rates. Such a behavior was found to be consistent with
investigations of the two flavor Schwinger model
\cite{PMV_Schwinger} for a quantity that is expected to vanish in the
chiral limit. There it was argued that the fast decay rare is
due to fluctuations within a given topological sector while the slow
decay rate is due to the presence of topology changing configurations.
Therefore, for $m_f=0$ one could try to fit the largest three $L_s$ points
to the form:
\begin{equation}
e_0 \exp(-e_1 L_s) \ .
\label{no_const__exponential}
\end{equation}
The results of the fit for the $8^4$, $\beta=2.3$, $m_f=0$, and
$L_s=16$, $20$, $24$ points as well as for the $4^4$, $\beta=2.1$,
$m_f=0$, and $L_s=24$, $32$, $40$ points are shown in table
\ref{tab_no_const_exponential}.
The fit to the $8^4$ data is acceptable.
However, the fit to the $4^4$ data has a
rather large $\cdof$. Because this fit is for larger $L_s$ than the
$8^4$ fit one would expect that if there were a second exponential
decaying to zero its effect would be more pronounced in the $4^4$ fit.
Therefore, the large $\cdof$ of the $4^4$ fit suggests that the
presence of a second exponential decaying to zero is not likely.
This could be made more precise if simulations with larger $L_s$ values
for the $8^4$ and $4^4$ lattices were done. However,
such simulations are beyond the computing resources of this
project. Also, the analysis in \cite{Shamir_weak_coupling} 
suggests functional
forms with more parameters. It would be interesting to fit to these
forms but that would require more data points and better statistics
both of which are also beyond the computing resources of this project.

Finally, the SUSY breaking by the irrelevant terms may have
non-negligible effects at the lattice spacings studied here. Although
it was found that the chiral condensate is non--zero at the chiral limit
in two lattice spacings, this is not enough to estimate its value in
the continuum limit.

\section{Conclusions}
\label{sec_conclusions}


The formulation of ${\cal N}=1$, SU(2) Super Yang-Mills theory on the lattice
with domain wall fermions (DWF) has several advantages over more traditional
lattice fermion regulators. Even at non-zero lattice spacing the chiral
limit can be taken by letting $L_s \to \infty$, where $L_s$ is
the number of sites along the fifth auxiliary direction. Since in that
limit there is no gluino mass term, supersymmetry is broken only by
irrelevant operators and there is no need for fine tuning.
Also, in that limit
the theory has exact zero modes on non-trivial topological
backgrounds.

However, even at finite $L_s$, where numerical simulations are done,
these properties are maintained to a good degree allowing
extrapolations to the $L_s \to \infty$ limit. Furthermore, the
Pfaffian resulting from the integration of Majorana fermions is
positive definite at finite $L_s$, non-zero lattice spacing and for
any background gauge field configuration. As a result, one can
unambiguously interpret it as a probability measure to be used by the
numerical simulation for importance sampling. 
This property also
allows the use of standard numerical algorithms where any number of
flavors $N_f$ can be simulated.
By contrast, Wilson fermions have this positivity property
only at the continuum limit.

In this work, the full ${\cal N}=1$, SU(2) Super Yang-Mills theory was
numerically simulated on the lattice using DWF. The gluino
condensate $\chibarchi$ was measured. These simulations did not
present any unexpected technical difficulties. 

A finite value of $L_s$ breaks chiral symmetry and induces a small
gluino mass. In addition, an explicit gluino mass $m_f$ was used to
provide extra control.  Several $m_f$ and $L_s$ values were used (all
corresponding to positive gluino mass) and the value of $\chibarchi$ 
was extrapolated to the chiral limit using three
different methods. All methods gave consistent results indicating small
systematic effects and
suggesting that the functions used for the fits
are consistent.
These
simulations were done on a lattice with $8^4$ lattice sites.

Additional simulations on a lattice with $4^4$ lattice sites but
approximately double the lattice spacing were done. Again,
extrapolations to the chiral limit gave a non-zero $\chibarchi$. In
this lattice $\left[\rm{mass} \times \rm{volume} \times \chibarchi \right] < 1$. 
Then analytical considerations suggest that the value of
$\chibarchi$ must come mostly from topological sectors with
fractional topological charge of $\pm 1/2$. Indeed, as the mass was
made smaller unusually large values (spikes) were observed in the
statistical sample of $\chibarchim$ indicating the singular
contribution of these sectors.

The spectrum of the theory is of great interest but it was not
possible to measure on the small lattices considered here.  Also, the
gluino condensate was measured only on two different lattice spacings
and therefore it was not possible to extrapolate to the continuum
limit where comparisons with analytical results would be
possible. Future work could explore these very interesting topics.


\section*{Acknowledgments}

This research was supported in part by NSF under grant \#
NSF-PHY96-05199 (J. Kogut and P. Vranas). All numerical simulations
were done on an 18 GFlops part of the QCDSP supercomputer at Columbia
University and on the 6 GFlops QCDSP supercomputer at Ohio State
University. We would like to thank N. Christ and R. Mawhinney for
providing us with the Columbia University resource and G. Kilcup for
providing us with the Ohio State University resource. Also, we would
like to thank Y. Shamir and T. Schaefer for useful
comments. P. Vranas
would like to thank D.B. Kaplan, A. Kovner, A. Nyffeler, and
E. Weinberg for enlightening discussions.


\section{Appendix}
\label{sec_appendix}

In this appendix the various tables are presented.
\vfill
\eject

\null
\begin{table}
\centering
\begin{tabular}{||c|c|c||} \hline
$L_s$	&	$m_f$	&	$\chibarchi$	 \\ \hline \hline
12      &       0.00    &       0.00902(4)       \\ \hline
12      &       0.02    &       0.01052(4)       \\ \hline
12      &       0.04    &       0.01223(5)       \\ \hline
12      &       0.06    &       0.01370(4)       \\ \hline
12      &       0.08    &       0.01519(3)       \\ \hline\hline
16      &       0.00    &       0.00694(7)       \\ \hline
16      &       0.02    &       0.00863(5)       \\ \hline
16      &       0.04    &       0.01026(4)       \\ \hline
16      &       0.06    &       0.01183(4)       \\ \hline
16      &       0.08    &       0.01324(4)       \\ \hline\hline
20      &       0.00    &       0.00588(5)       \\ \hline
20      &       0.02    &       0.00735(10)      \\ \hline
20      &       0.04    &       0.00897(7)       \\ \hline
20      &       0.06    &       0.01071(3)       \\ \hline
20      &       0.08    &       0.01221(3)       \\ \hline\hline
24      &       0.00    &       0.00516(6)       \\ \hline
24      &       0.02    &       0.00691(4)       \\ \hline
24      &       0.04    &       0.00827(7)       \\ \hline
24      &       0.06    &       0.00992(3)       \\ \hline
24      &       0.08    &       0.01142(3)       \\ \hline
\end{tabular}
\caption{The values of $\chibarchi$ for the $8^4$ simulations 
at $\beta=2.3$, $m_0=1.9$.} 
\label{tab_8888_data}
\end{table}

\null
\begin{table}
\centering
\begin{tabular}{||c|c|c||} \hline
$L_s$	&	$m_f$	&	$\chibarchi$	  \\ \hline \hline
16      &       0.00    &        0.00743(14)      \\ \hline
24      &       0.00    &        0.00474(10)      \\ \hline
32      &       0.00    &        0.00351(7)       \\ \hline
40      &       0.00    &        0.00308(11)      \\ \hline
\end{tabular}
\caption{The values of $\chibarchi$ for the $4^4$ simulations 
at $\beta=2.1$, $m_0=1.9$.} 
\label{tab_4444_data}
\end{table}

\null
\begin{table}
\centering
\begin{tabular}{||c|c|c|c||} \hline
Figure	&	$b_0$		&	$b_1$		&	$\cdof$	\\ \hline \hline
3.a	&       0.00904(5)      &       0.0772(8)       &       3.6     \\ \hline
3.b	&       0.00717(6)      &       0.0767(10)      &       3.8     \\ \hline
3.c	&       0.00585(9)      &       0.0799(13)      &       3.5     \\ \hline
3.d	&       0.00538(5)      &       0.0755(8)       &       2.2     \\ \hline \hline
6	&       0.00455(21)     &       0.0704(37)      &       2.6     \\ \hline \hline
\end{tabular}
\caption{The results of the linear fits 
presented in the various figures to the function $b_0 + b_1 m_f$ 
(dof is short for degree of freedom).}
\label{tab_linear_fits}
\end{table}

\null
\begin{table}
\centering
\begin{tabular}{||c|c|c|c|c||} \hline
Figure	&	$c_0$		&	$c_1$		&	$c_2$		&	$\cdof$	\\ \hline\hline
4	&       0.00444(21)     &       0.023(3)        &       0.135(13)	&	3.9	\\ \hline\hline
5.a	&	0.01034(16)     &       0.021(1)        &       0.123(8)        &       6.2	\\ \hline
5.b	&	0.00857(19)     &       0.019(1)        &       0.111(8)        &       1.1	\\ \hline
5.c	&	0.00700(25)     &       0.022(2)        &       0.119(10)       &       0.6	\\ \hline
5.d	&	0.00611(16)     &       0.025(3)        &       0.143(12)       &       2.9	\\ \hline\hline
7	&	0.00432(22)     &       0.025(3)        &       0.141(13)       &       1.4	\\ \hline\hline
8	&	0.00268(19)     &       0.026(4)        &       0.107(12)       &       0.4	\\ \hline\hline
\end{tabular}
\caption{The results of the exponential fits 
presented in the various figures to the function $c_0 + c_1 \exp(-c_2 L_s)$
(dof is short for degree of freedom).}
\label{tab_exp_fits}
\end{table}

\null
\begin{table}
\centering
\begin{tabular}{||c|c||} \hline
Method	&	$\chibarchi(m_f \to 0,L_s \to \infty)$	 \\ \hline \hline
I	&	0.00444(21)	\\ \hline
II	&	0.00455(21)	\\ \hline
III	&	0.00432(22)	\\ \hline
\end{tabular}
\caption{The $m_f \to 0$, $L_s \to \infty$ extrapolated values of $\chibarchi$ using
the three different extrapolation methods described in the text.}
\label{tab_extrap}
\end{table}

\null
\begin{table}
\centering
\begin{tabular}{||c|c|c|c|c||} \hline
Figure	&	$d_0$		&	$d_1$		&	$d_2$		&	$\cdof$	\\ \hline\hline
7	&	 0.000(2)       &       0.063(19)       &       -0.78(20)       &       11	\\ \hline
8	&	-0.016(21)	&       0.043(12)       &       -0.23(24)	&       35	\\ \hline
\end{tabular}
\caption{The results of the power law fits in the data of figures 7 and 8
to the function $d_0 + d_1 L_s^{d_2}$ (dof is short for degree of
freedom).}
\label{tab_plaw_fits}
\end{table}

\null
\begin{table}
\centering
\begin{tabular}{||c|c|c|c|c|c|c||} \hline
Lattice&$\beta$ & $m_f$	& $L_s$		&	$e_0$	&	$e_1$	&	$\cdof$ \\ \hline\hline
$8^4$	& 2.3	& 0.0	& 16,20,24 	&  0.0125(5)    & 0.037(2)	&	2.8	\\ \hline\hline
$4^4$	& 2.1	& 0.0	& 24,32,40	&  0.0094(7)    & 0.030(2)	&	9.1	\\ \hline
\end{tabular}
\caption{The results of the exponential fits without a constant  using
the function $e_0 \exp(-e_1 L_s) $
(dof is short for degree of freedom).}
\label{tab_no_const_exponential}
\end{table}

\vfill
\eject

%
%
\if \epsfpreprint Y \eject

\figure{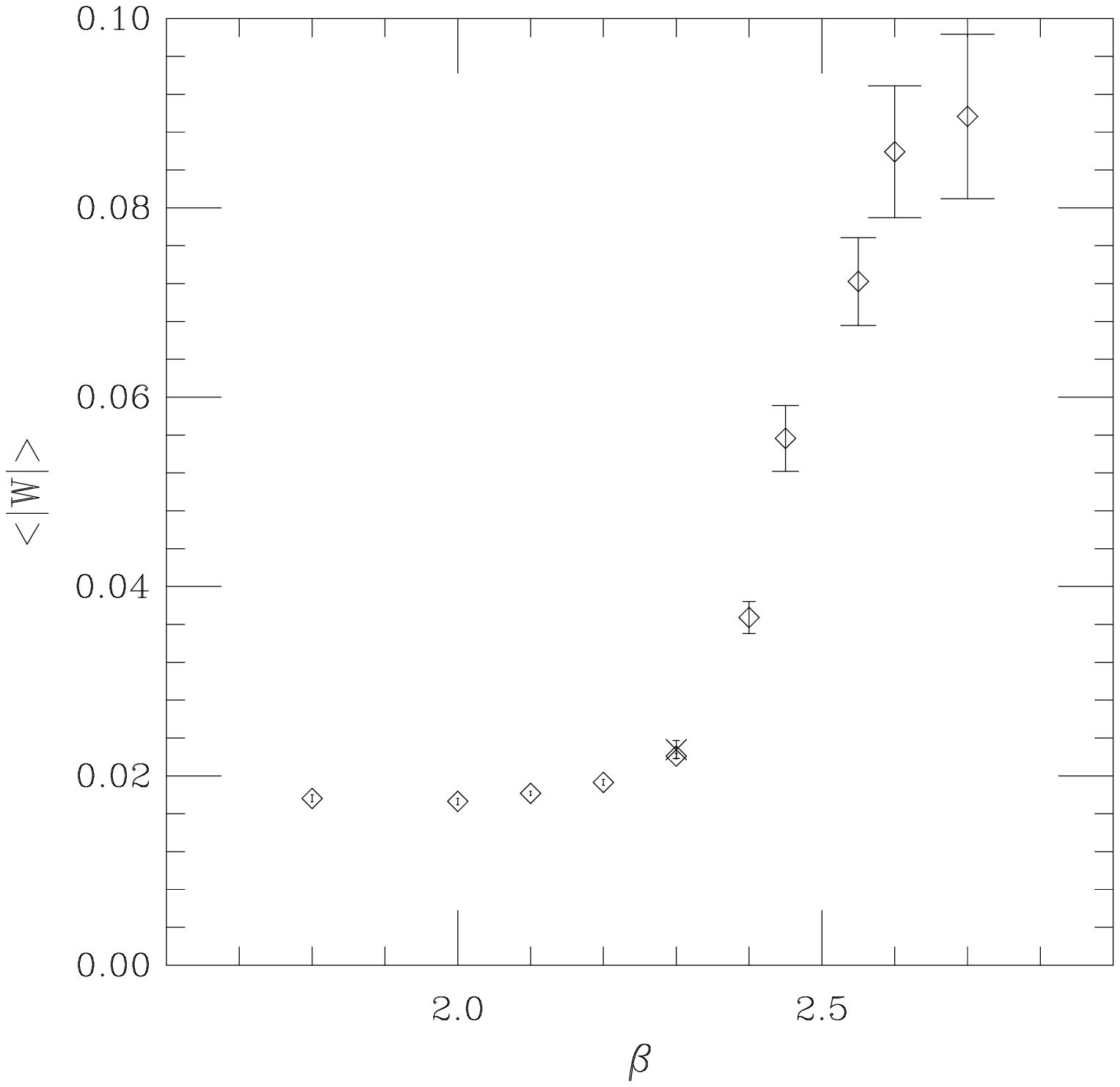}{\one}{\figsizea} \eject

\figure{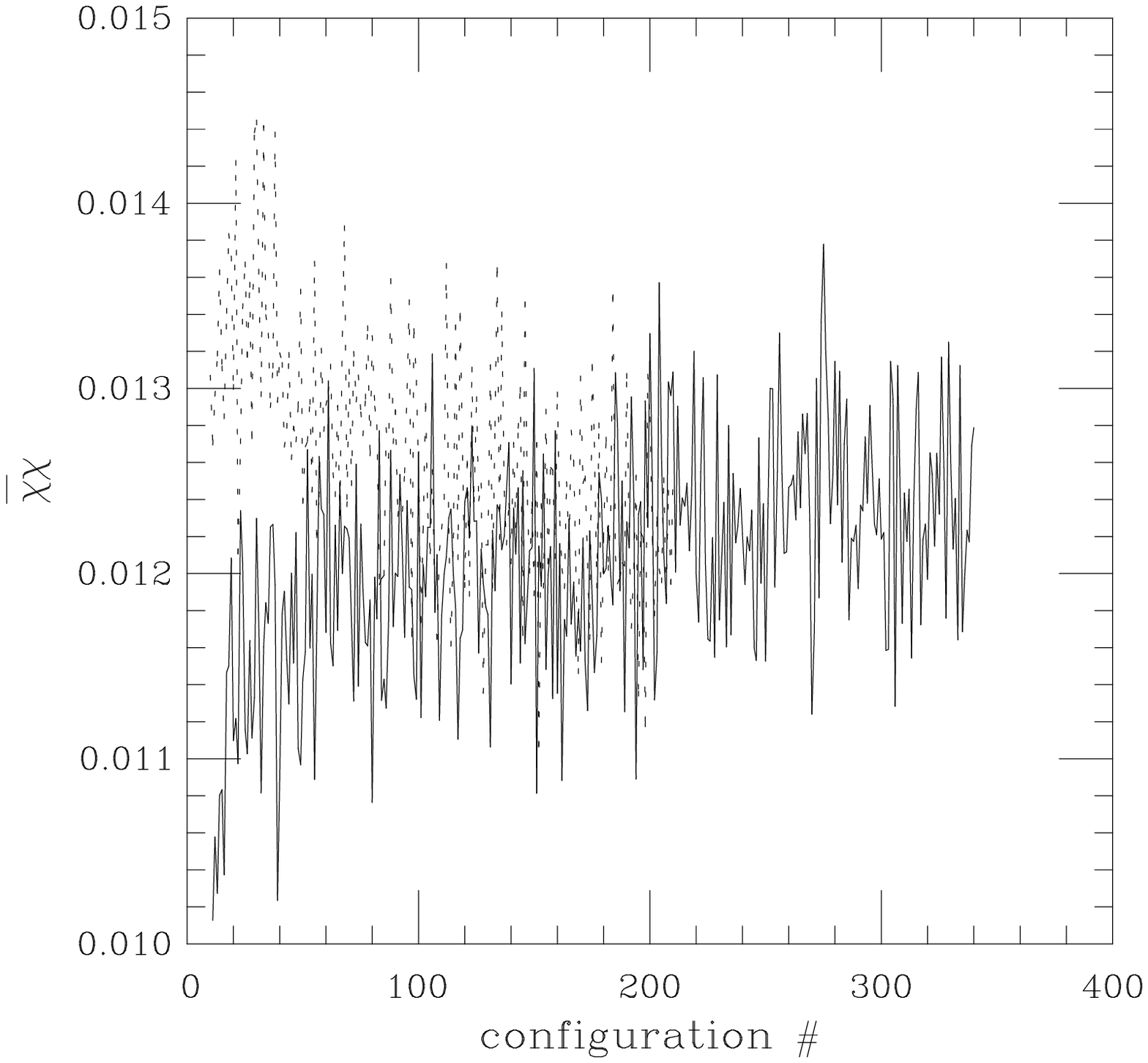}{\two}{\figsizea} \eject

\figure{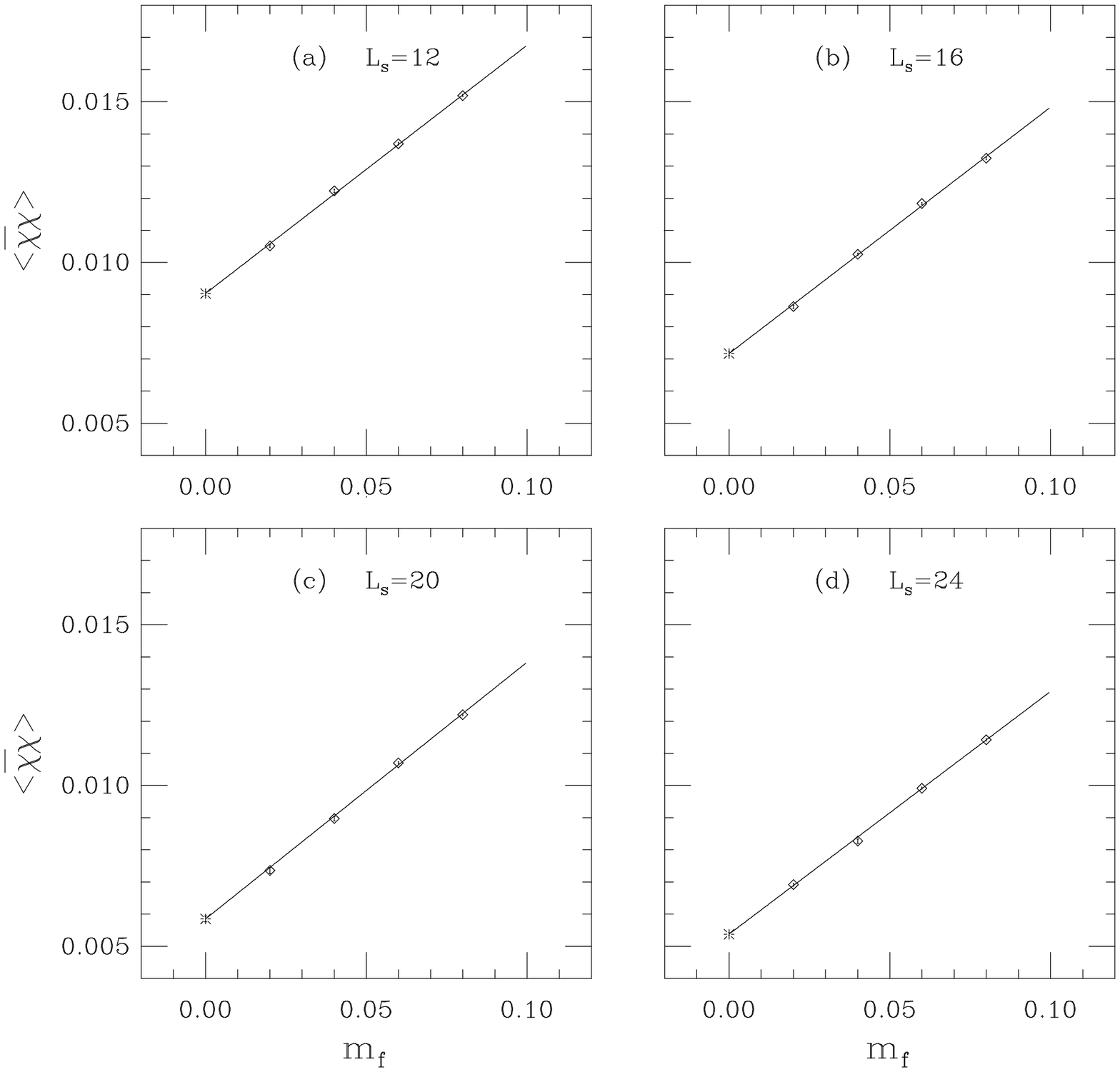}{\three}{\figsizea} \eject

\figure{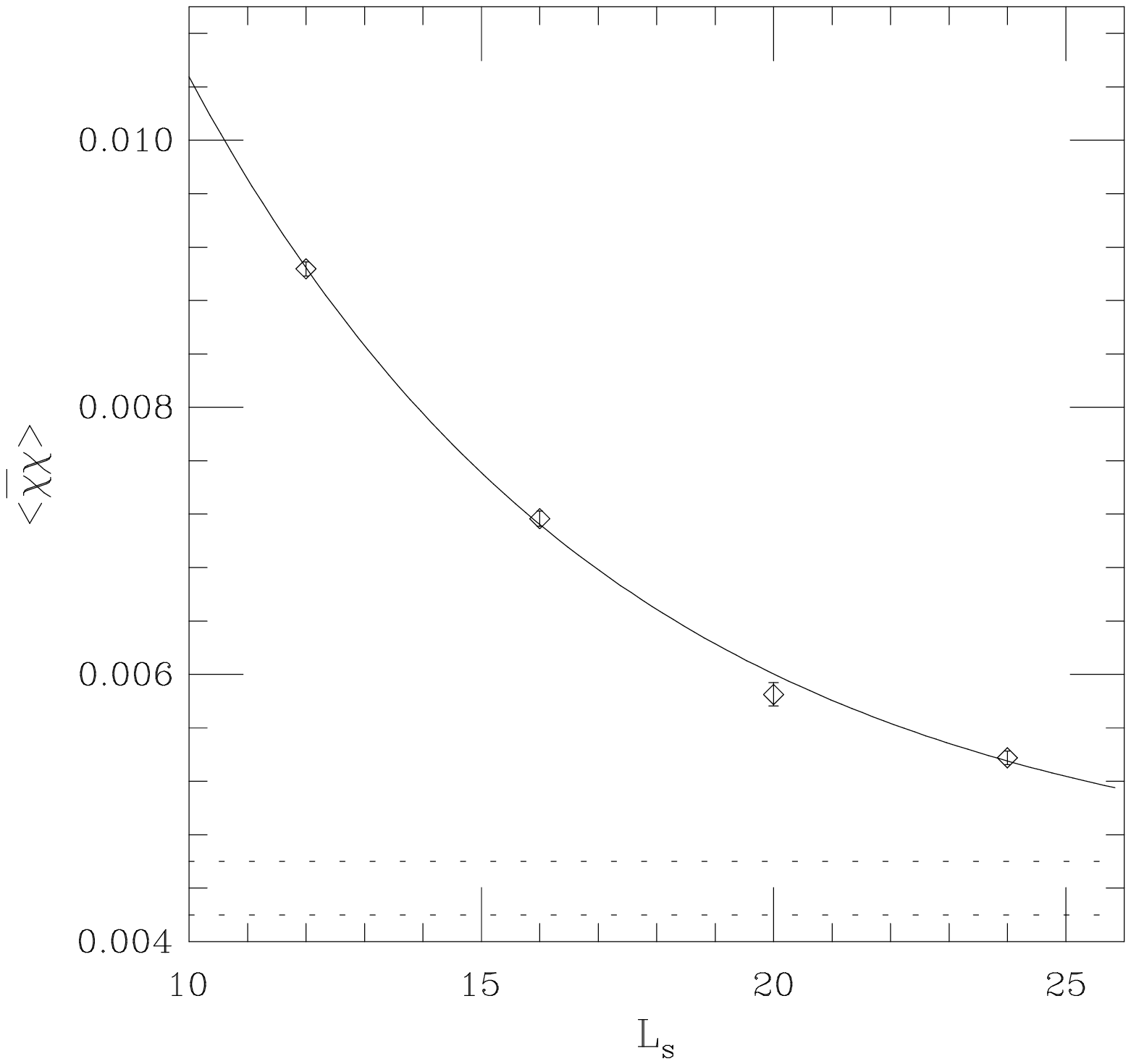}{\four}{\figsizea} \eject

\figure{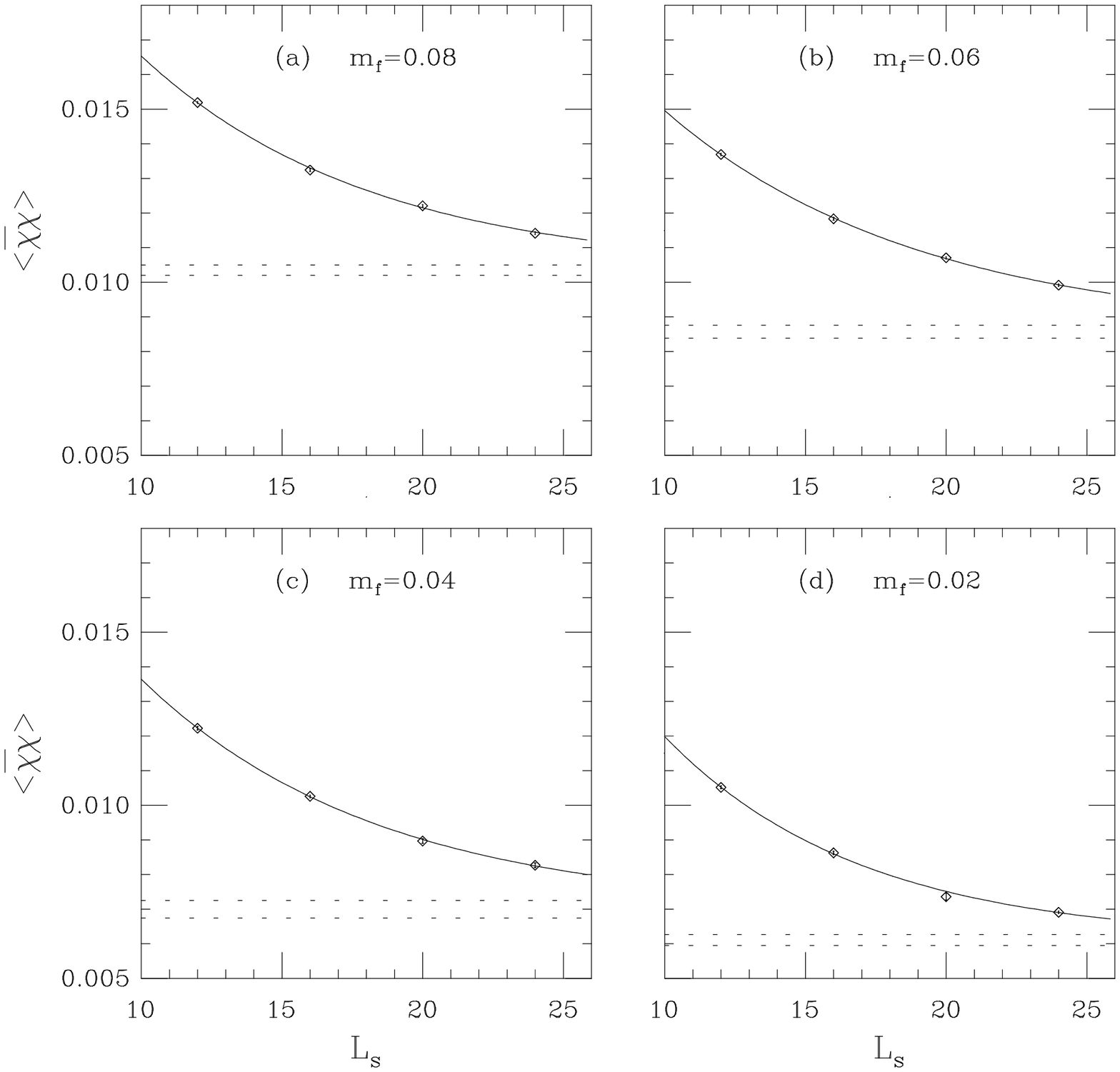}{\five}{\figsizea} \eject

\figure{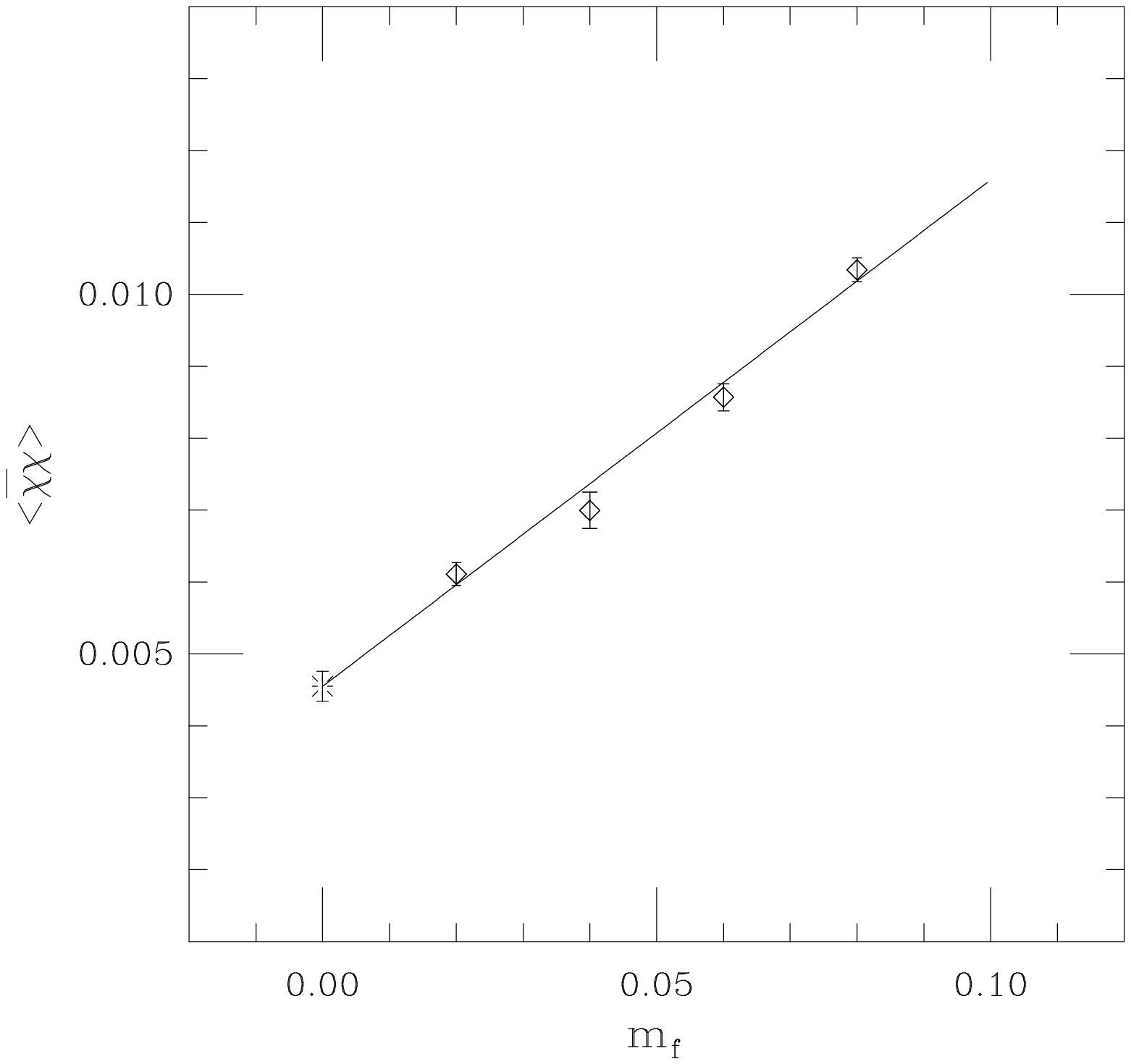}{\six}{\figsizea} \eject

\figure{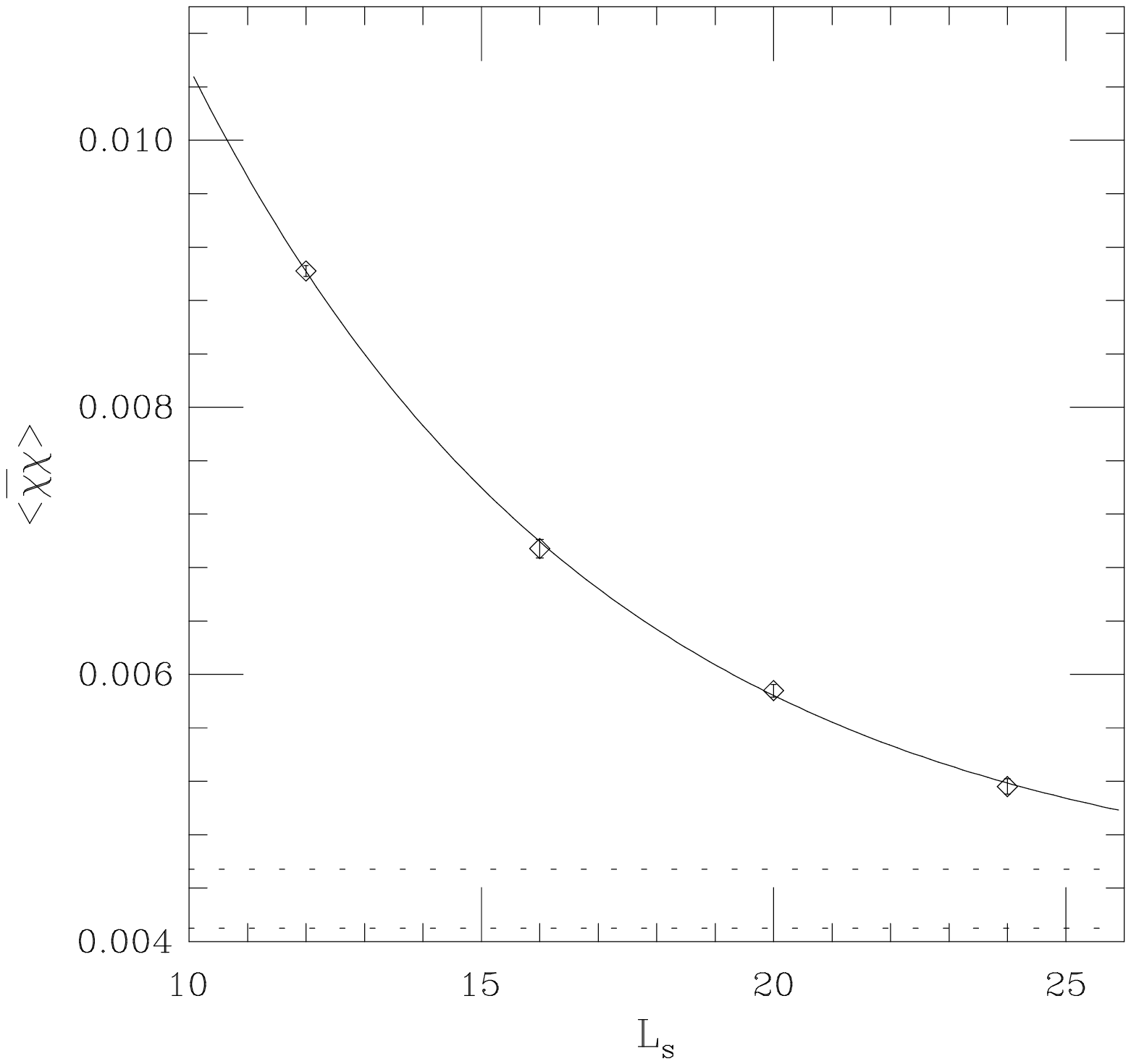}{\seven}{\figsizea} \eject

\figure{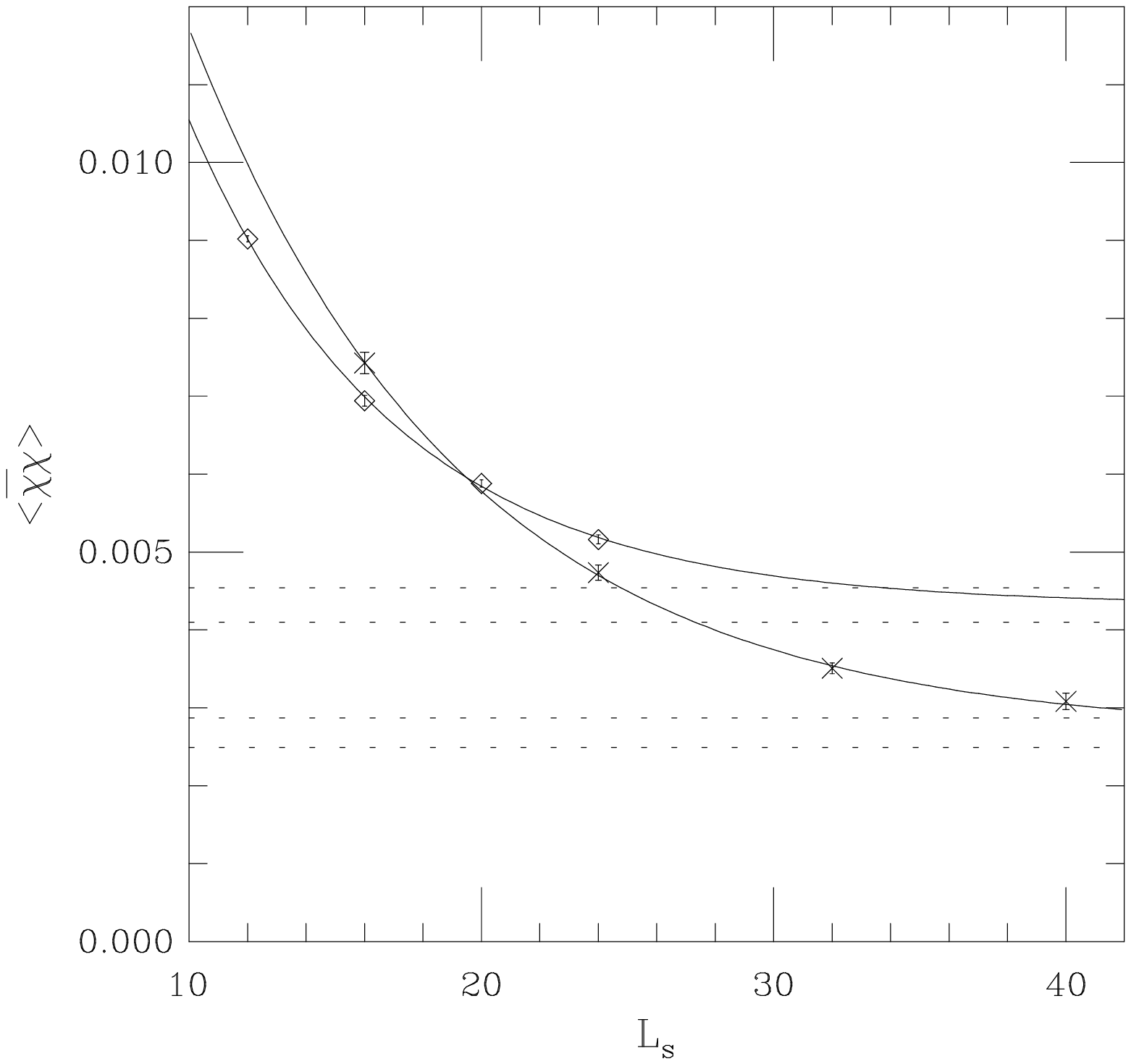}{\eight}{\figsizea} \eject

\figure{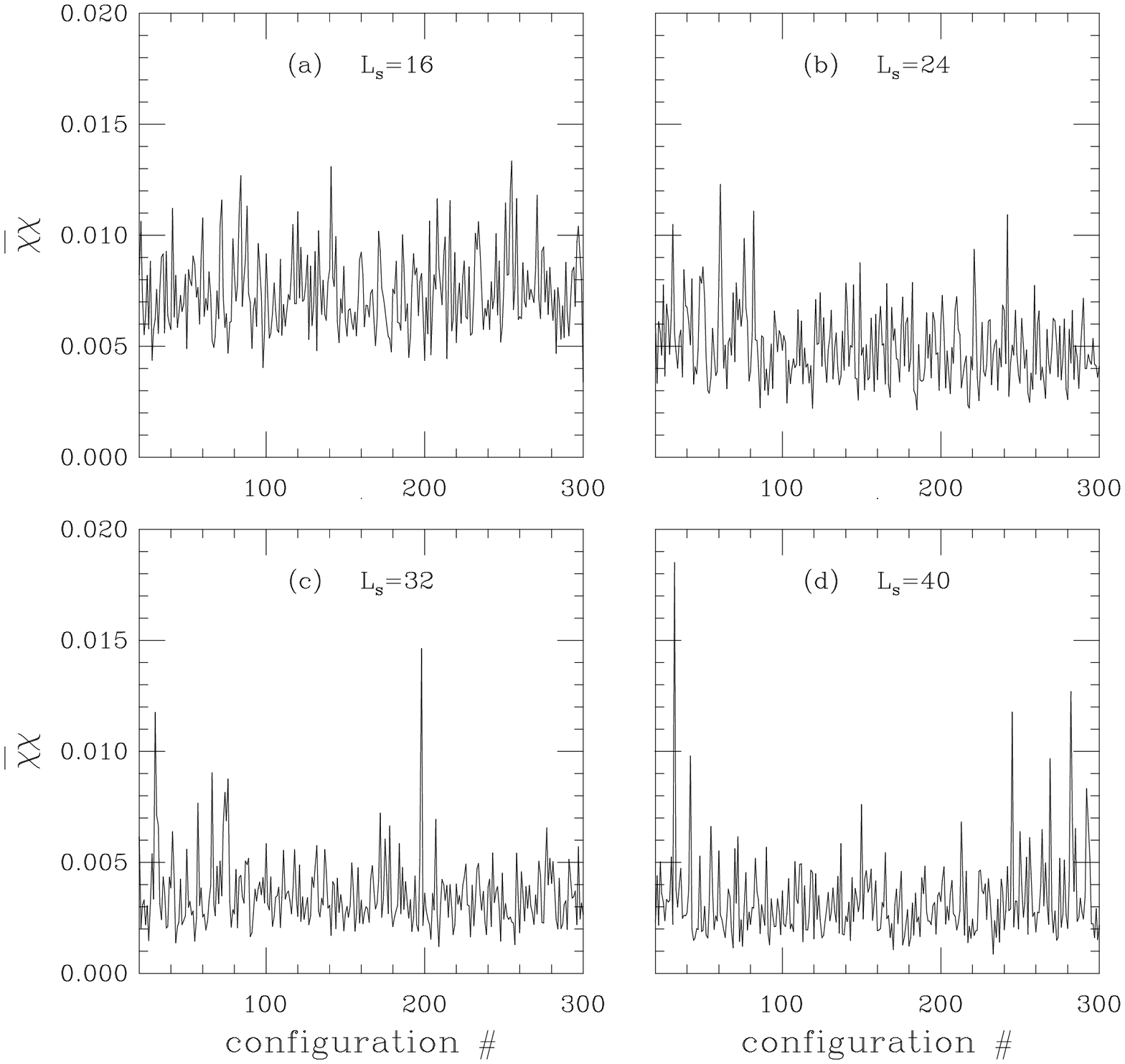}{\nine}{\figsizea} \eject

\figure{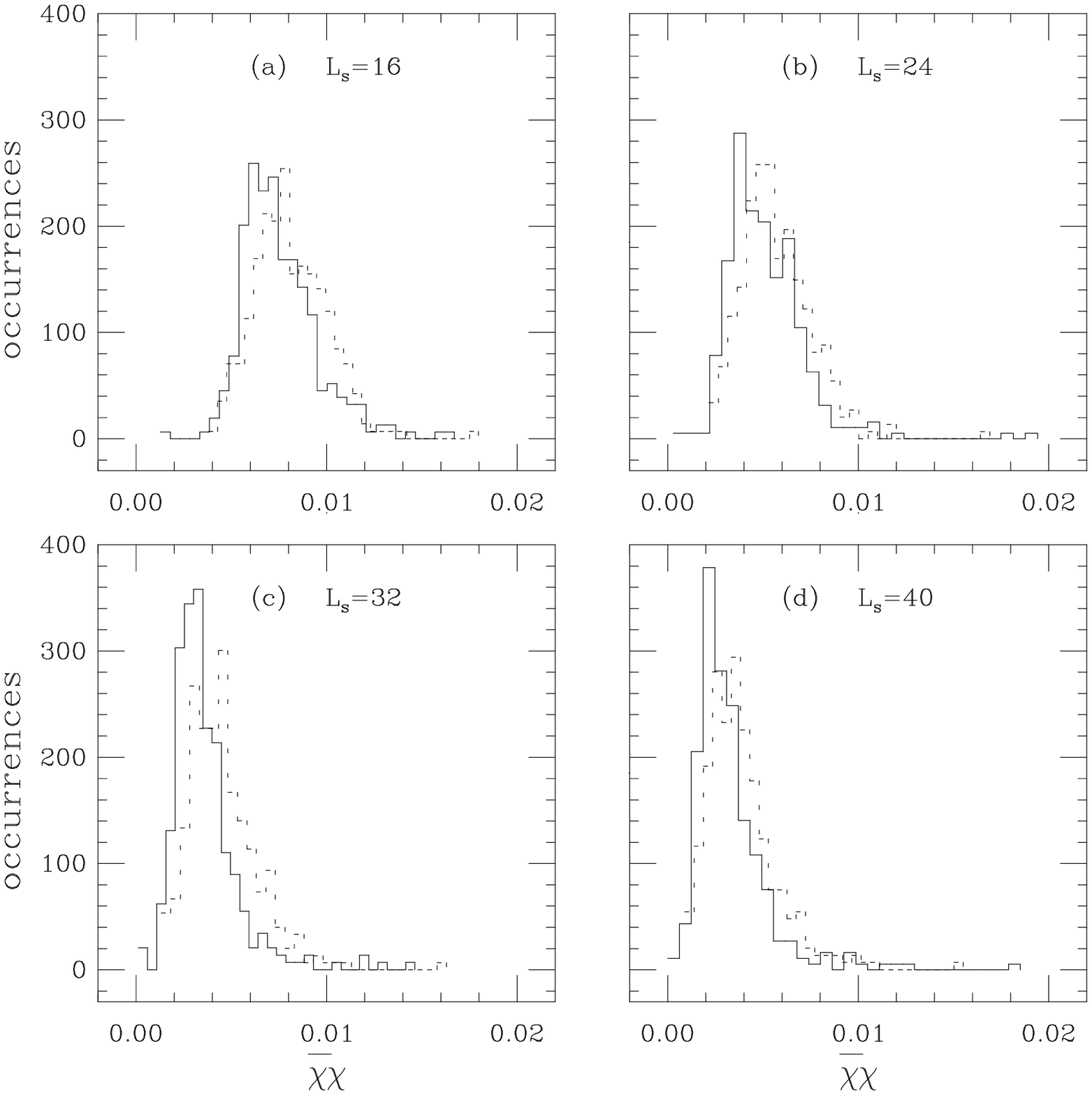}{\ten}{\figsizea} \eject

\figure{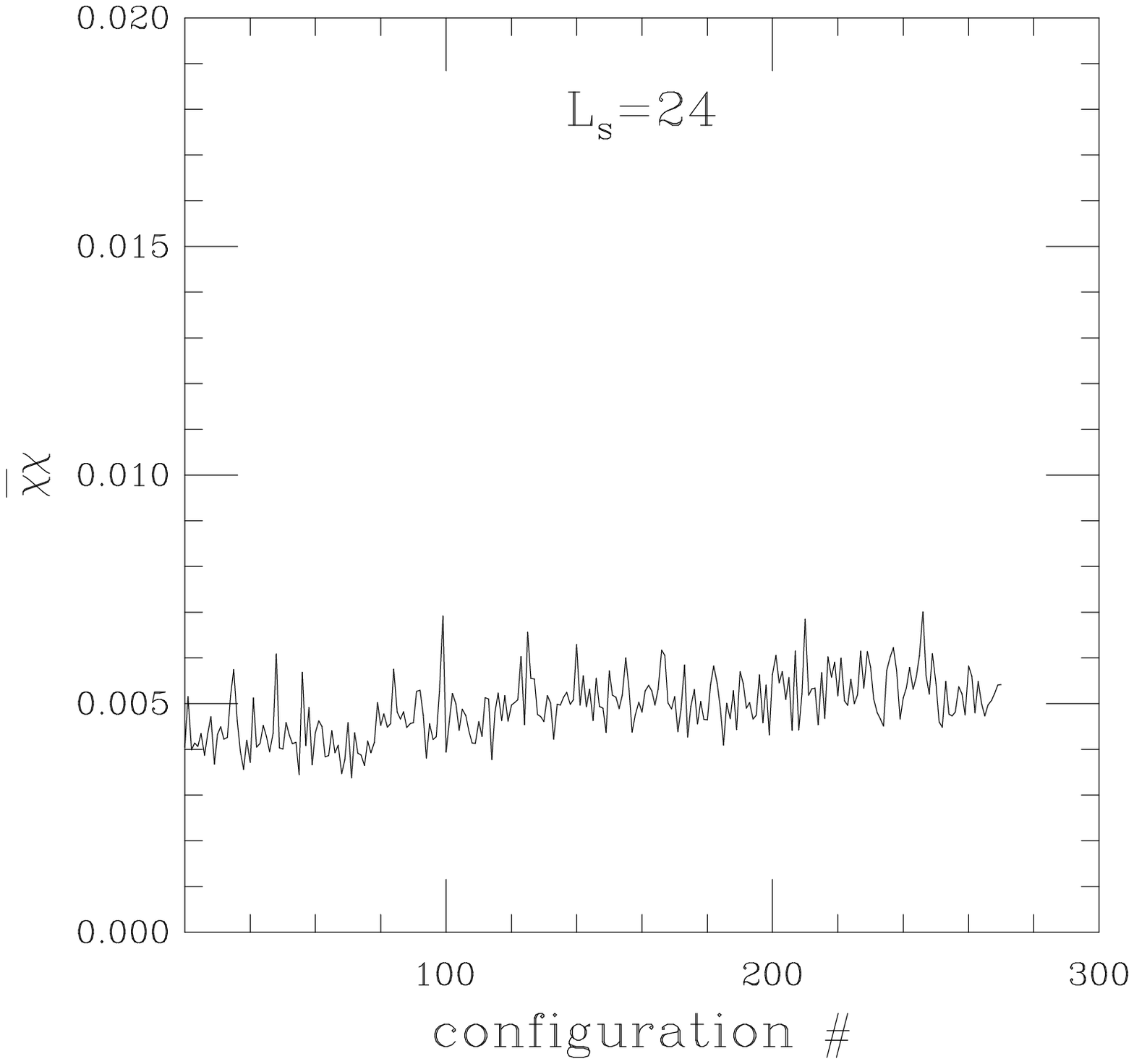}{\eleven}{\figsizea} \eject

\fi 
%
\if \epsfpreprint N \eject 
\section* {Figure Captions.}
\noindent{\bf Figure 1:} \one 
\noindent{\bf Figure 2:} \two 
\noindent{\bf Figure 3:} \three 
\noindent{\bf Figure 4:} \four 
\noindent{\bf Figure 5:} \five
\noindent{\bf Figure 6:} \six
\noindent{\bf Figure 7:} \seven
\noindent{\bf Figure 8:} \eight
\noindent{\bf Figure 9:} \nine
\noindent{\bf Figure 10:} \ten
\fi
\end{document}